\documentclass[lettersize,journal]{IEEEtran}
\usepackage{amsmath,amsfonts}
\usepackage{algorithmic}
\usepackage{algorithm}
\usepackage{array}
\usepackage[caption=false,font=normalsize,labelfont=sf,textfont=sf]{subfig}
\usepackage{textcomp}
\usepackage{stfloats}
\usepackage{url}
\usepackage{verbatim}
\usepackage{graphicx}
\usepackage{cite}
\usepackage{courier}
\input{insbox.tex}
\begin{document}

\title{Efficient Computation of Voronoi Diagrams \\ Using Point-in-Cell Tests}

\author{Yanyang Xiao, Juan Cao, and Zhonggui Chen
\thanks{Manuscript received February 16, 2026. The work of Yanyang Xiao was supported by the National Natural Science Foundation of China (No. 62562046) and Jiangxi Provincial Natural Science Foundation, China (Nos. 20232BAB212010, 20242BAB25072). The work of Juan Cao and Zhonggui Chen was supported by National Natural Science Foundation of China (Nos. 62272402, 62372389), and Natural Science Foundation of Fujian Province, China (No. 2024J01513243). (\textit{Corresponding author: Zhonggui Chen.})}
\thanks{Yanyang Xiao is with the School of Mathematics and Computer Sciences, Nanchang University, Nanchang 330031, China (e-mail: xiaoyanyang@ncu.edu.cn).}
\thanks{Juan Cao is with the School of Mathematical Sciences, Xiamen University, Xiamen, 361005, China (e-mail: juancao@xmu.edu.cn).}
\thanks{Zhonggui Chen is with the School of Informatics, Xiamen University, Xiamen, 361005, China (e-mail: chenzhonggui@xmu.edu.cn).}
}

\markboth{IEEE Transactions on Visualization and Computer Graphics,~Vol.~14, No.~8, August~2021}%
{Xiao \MakeLowercase{\textit{et al.}}: Efficient Computation of Voronoi Diagrams Driven by Point-Cell Relationship}


\maketitle

\begin{abstract}
Since the Voronoi diagram appears in many applications, the topic of improving its computational efficiency remains attractive. We propose a novel yet efficient method to compute Voronoi diagrams bounded by a given domain, i.e., the clipped or restricted Voronoi diagrams. The intersection of the domain and a Voronoi cell (domain-cell intersection) is generated by removing the part outside the cell from the domain, which can be accomplished by several clippings. Different from the existing methods, we present an edge-based search scheme to find clipping planes (bisectors). A test called point-in-cell is first set up to tell whether a space point is in a target Voronoi cell or not. Then, for each edge of the intermediate domain-cell intersection, we will launch a clipping only if its two endpoints are respectively inside and outside the corresponding Voronoi cell, where the bisector for the clipping can be found by using a few times of point-in-cell tests. Therefore, our method only involves the clippings that contribute to the final results, which is a great advantage over the state-of-the-art methods. Additionally, because each domain-cell intersection can be generated independently, we extend the proposed method to the GPUs for computing Voronoi diagrams in parallel. The experimental results show the best performance of our method compared to state-of-the-art ones, regardless of site distribution. This paper was submitted to IEEE Transactions on Visualization and Computer Graphics in February 2026. 
\end{abstract}


\section{Introduction}
\IEEEPARstart{G}{iven} a set of points (called sites or generators), a Voronoi diagram (VD)~\cite{Aurenhammer1991-Voronoi} decomposes the space into small cells (called Voronoi cells). Each cell consists of points whose distance to the corresponding site is not greater than that to the other sites, where the distance metric is usually the Euclidean distance. The Voronoi diagram is frequently used in a wide range of applications, including mesh generation~\cite{Du1999-CVT,Yan2013-ClippedVoronoi,Jie2014-FEMMesh,Cao2022-FEMMesh}, surface remeshing~\cite{Yan2009-Remeshing,Levy2010-LpCVT,Levy2013-kNNVoronoi,Nivoliers2015-Remeshing}, sampling~\cite{Balzer2009-CCVT,Chen2012-Sampling,Yan2015-Sampling}, image processing~\cite{Chen2014-VoroApprox,Duan2015-ImagePolygons,Cao2018-VoroApprox}, point cloud processing~\cite{Boltcheva2018-Reconstruction,Chen2018-CVTResampling,Xiao2025-Resampling}, etc. Improving the computational efficiency of Voronoi diagrams remains attractive, especially for those applications requiring iterative construction of Voronoi diagrams.

For ordinary Voronoi diagrams, some Voronoi cells are unbounded, containing infinitely distant points in the space. In practice, given a domain composed of a set of simplices (triangles in the 2D and surface cases or tetrahedra in the 3D case), most applications are interested in the intersections of the Voronoi cells and the domain (domain-cell intersections), which are called clipped or restricted Voronoi cells. Thus, our goal is to compute domain-cell intersections as fast as possible, each of which consists of simplex-cell intersections.

The computation of Voronoi diagrams has been well studied in recent decades. Although many methods have been presented, each of them is with inherent defects. Some algorithms including divide-and-conquer~\cite{Shamos1975-DivideAndConquerVoronoi}, incremental insertion~\cite{Green1978-DirichletIncrement}, and sweep-line~\cite{Fortune1987-SweepVoronoi} were developed only for the planar domains and are hard to use in higher dimensional spaces. Despite jump flooding algorithm (JFA)-based method~\cite{Rong2006-JFAVoronoi,Rong2011-JFACVT} being capable of dealing with 2D and 3D cases, it was designed to cluster pixels from a preset texture for parallel computation on GPUs, thus generating approximate Voronoi cells and being incompatible with complex domains. To obtain Voronoi diagrams of general cases, the Delaunay triangulation (DT)-based~\cite{Aurenhammer1991-Voronoi, Yan2009-Remeshing,Yan2013-ClippedVoronoi} and \textit{k}-nearest neighbors (\textit{k}NN)-based methods~\cite{Levy2013-kNNVoronoi} are more popular. However, the former highly relies on the Delaunay triangulation of the given sites, while the construction of triangulation is time-consuming. The latter which solely depends on \textit{k}NN queries has attracted much attention in recent years. The cell computation is independent of each other and suitable for parallel computation~\cite{Ray2018-MeshlessVoronoi,Han2017-GPURVD,Liu2022-GPU3DVoronoi,Basselin2021-GPUknnRPD}. The main drawback of \textit{k}NN-based method is that its efficiency is sensitive to the distribution of the given sites~\cite{Ray2018-MeshlessVoronoi}. In high dimensional spaces, the \textit{k}NN-based method usually performs worse than the DT-based method when the distribution leads to long and thin Voronoi cells. Another method called corner validation~\cite{Sainlot2017-CornerValidation} shows better performance than the \textit{k}NN-based method when the sites are far away from the domain. However, if a simplex intersects with a large number of Voronoi cells, the corner validation method requires more iterations, resulting in a significant computational cost.

\begin{figure*}[!t]
\centering
\subfloat[white noise]{\includegraphics[height=0.24\textwidth]{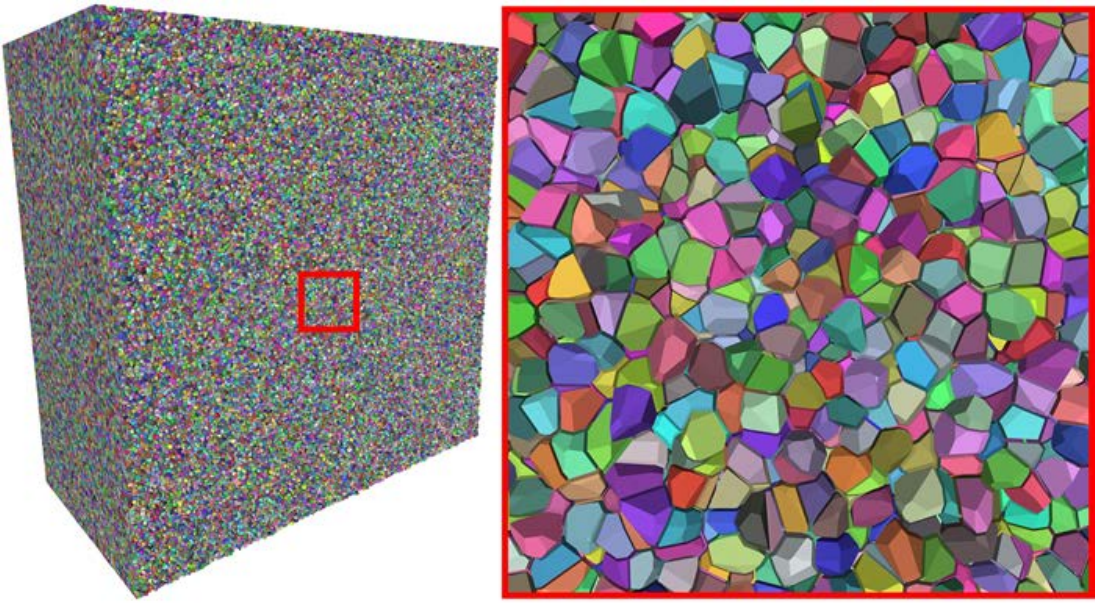}} \quad \quad \quad
\subfloat[blue noise]{\includegraphics[height=0.24\textwidth]{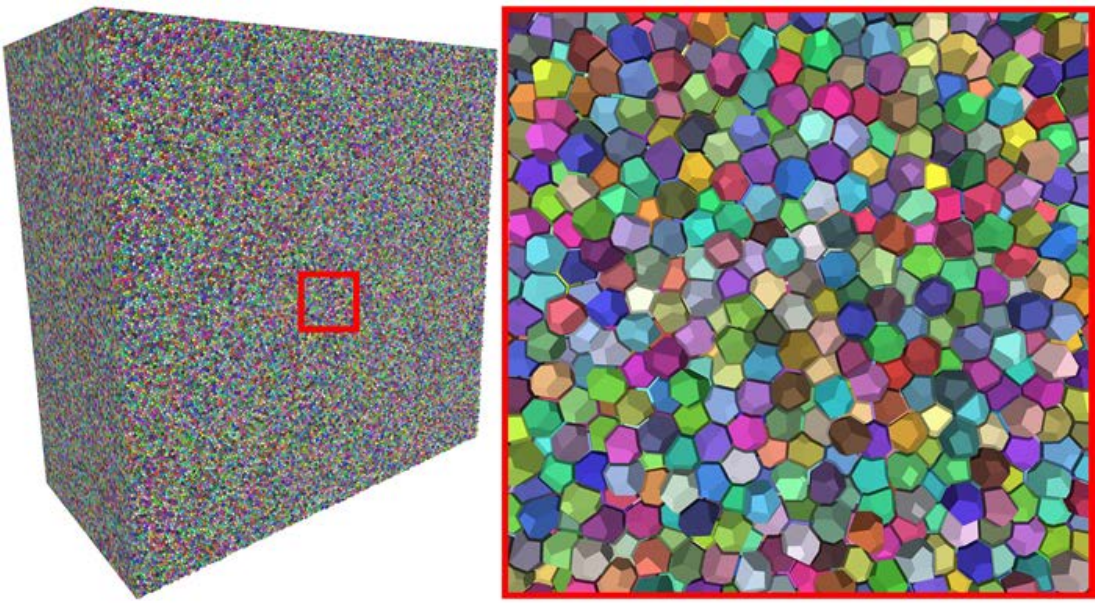}}
\caption{Computing 5M 3D Voronoi cells bounded by a cube, where the given sites are white noise (a) and blue noise (b), respectively. Tested on an NVIDIA RTX 3080 GPU, the runtimes of two parallel implementations~\cite{Ray2018-MeshlessVoronoi, Basselin2021-GPUknnRPD} of the \textit{k}NN-based method and our method are (a): 3.12, 2.81, \textbf{0.49} seconds and (b): 0.34, 0.42, \textbf{0.26} seconds, respectively. Only a subset of the cells are rendered, each shrunk for visualization.}
\label{fig:teaser}
\end{figure*}

We propose an efficient method to compute clipped or restricted Voronoi diagrams, where the computation of simplex-cell intersections is fundamentally distinct from existing methods. We first set up a point-in-cell test to tell whether a space point is in a target Voronoi cell or not. When computing a simplex-cell intersection, at most one clipping will be launched for each edge of the intersection, where the bisector for the clipping can be found by using a few times of the point-in-cell tests. Thus, our method only involves valid clippings to the final results, which is a great advantage over the existing methods, including DT-based, \textit{k}NN-based, and corner validation, since a part of their clippings are useless to results but will still be executed. This is the key to our method in obtaining higher efficiency than existing methods. The main contributions of this paper can be summarized as:

\begin{itemize}
    \item A novel solution for generating simplex-cell intersections is presented. For each edge of an intersection, we launch at most one clipping that is guaranteed-valid to the result, where the corresponding bisector can be found quickly by using the point-in-cell tests.
    \item An efficient method for computing Voronoi cells bounded by a given domain is proposed. Extensive comparisons on 2D, surface, and 3D cases demonstrate that our method outperforms existing approaches, regardless of site distribution.
    \item By modifying the presented solution for simplex-cell intersection computation, the proposed method can be easily adapted to GPUs for computing clipped or restricted Voronoi diagrams in parallel. Our parallel algorithm is demonstrated to achieve higher efficiency compared to the state-of-the-art \textit{k}NN-based algorithm.
\end{itemize}

In the rest of this paper, Section \ref{sec:related-work} reviews related work on the computation of Voronoi diagrams, and Section \ref{sec:preliminaries} provides the necessary preliminaries. Our proposed method is introduced in Section \ref{sec:proposed-method}, followed by its parallel implementation in Section \ref{sec:parallel-computation}. Implementation details and experimental results are presented in Sections \ref{sec:implementation-details} and \ref{sec:results}, respectively. Finally, Section \ref{sec:conclusion} concludes the paper.

\section{Related Work}
\label{sec:related-work}

We focus on methods capable of computing both 2D and 3D Voronoi cells within a given domain. A simplex-cell intersection is typically generated by some clippings, while finding the corresponding bisectors varies across different methods. A clipping is contributed or valid only if it influences the correctness of the final result. We propose a novel solution for generating simplex-cell intersections that involves only valid clippings. This contrasts with state-of-the-art approaches, such as DT-based and \textit{k}NN-based methods, which may incorporate invalid ones.

\subsection{DT-based method}

It is known that Voronoi diagram is the dual graph of Delaunay triangulation, based on which, the Voronoi diagram can be computed by using its duality~\cite{Aurenhammer1991-Voronoi, Yan2009-Remeshing,Yan2013-ClippedVoronoi}. Once the Delaunay triangulation of the given sites is established, we can explicitly obtain incident neighbors of each site, so that the corresponding Voronoi cell can be generated by clipping the domain using the bisectors of the site and its neighbors. The key to this method is the construction of Delaunay triangulation. Although there have been many algorithms, including flipping~\cite{Lawson1972-Delaunay, Joe1989-FlipDelaunay}, incremental insertion~\cite{Bowyer1981,Watson1981,Guibas1985-IncrementalDelaunay,Guibas1992-IncrementalDelaunay}, and divide-and-conquer~\cite{Dwyer1987-DivideConquerDelaunay,Cignoni1998-DeWallDelaunay}, the time cost to build triangulation is still high for Voronoi computation.

Another issue of the DT-based method being easily ignored is the existence of invalid clippings during the computation of domain-cell intersections. Imagine the case that a Voronoi cell covers multiple simplices. To generate the simplex-cell intersection, only a part of the clippings related to the Voronoi cell are useful, while the rest are useless but will still be executed, causing efficiency loss. See Fig. \ref{fig:geometry}(b) as an example, only two clippings are valid for generating the intersection (light green), with the relevant bisectors originating from the red site and two purple sites.

The convex hull-based method~\cite{Brown1979-ConvexhullVoronoi} lifts the given sites to the space of one dimension higher and constructs a convex hull of these lifted points for computing Voronoi diagrams. It is essentially the same as the DT-based method, because the connectivity of the convex hull is identical to that of the Delaunay triangulation, thereby exhibiting similar drawbacks.


\subsection{\textit{k}NN-based method}

Each pair of sites divides the space into two halves, and all the half-spaces associated with a site intersect in a common area, namely its Voronoi cell. To obtain it, we can construct the bisectors of the site and all the others to clip the given domain. Apparently, it is unnecessary to consider all other sites, as a Voronoi cell is primarily defined by bisectors of its nearest neighbors. Therefore, the \textit{k}NN-based method~\cite{Levy2013-kNNVoronoi} finds \textit{k}-nearest neighbors of the site, where the neighbors are sorted in ascending order according to their distances to the site, then the bisectors of the site and its neighbors are sequentially used to clip the domain for generating the Voronoi cell. The cell computation completes once the security radius is found, i.e., the maximum distance from the site to its cell is smaller than the distance from the site to any of the rest bisectors. Benefiting from the independence of cell computation, the parallelization of this method on GPUs has been well studied~\cite{Ray2018-MeshlessVoronoi,Han2017-GPURVD,Liu2022-GPU3DVoronoi,Basselin2021-GPUknnRPD}. Moreover, this method can also be extended to compute power diagrams~\cite{Basselin2021-GPUknnRPD,deGoes2015-PowerPaticles, Xiao2023-knnpower}.

The \textit{k}NN-based method suffers from invalid clippings as well. If a site dominates points far away from the site, the method has to visit more neighbors to clip the domain, since the security radius of such a case is hard to meet. All these clippings will be executed even though only a few of them are valid, explaining that the method is time-consuming once the given site distribution does not exhibit blue noise properties. The example in Fig. \ref{fig:geometry}(b) shows that bisectors of the red site and gray sites are useless to generate the simplex-cell intersection (light green) if the \textit{k}NN-based method is applied. Although users can set a maximum number of neighbors to avoid excessive clippings, the resultant cells are possibly an approximation of the exact Voronoi cells if the number is not large enough.


\subsection{Parallel algorithms}

The parallel computation of Voronoi diagrams on GPUs and multi-cores CPUs has remained an active research focus. Given their duality, the parallel construction of Delaunay triangulations is likewise an area of ongoing interest (e.g., \cite{Cao2014-GPUDelaunay,Marot2019-ParallelDelaunay,Chen2012-ParallelDelaunay,Gao2025-GPUDelaunay3D}). Compared to traditional algorithms (e.g., \cite{Cao2014-GPUDelaunay,Marot2019-ParallelDelaunay}) that require data synchronization, the methods based on the local Delaunay lemma~\cite{Chen2012-ParallelDelaunay,Gao2025-GPUDelaunay3D} independently construct the one-ring neighboring Delaunay simplices for each site, thus enabling a higher degree of parallelism. Since the intersections of the given domain and Voronoi cells are central to computing Voronoi diagrams, we focus on parallel algorithms for directly generating clipped (or restricted) Voronoi cells. The JFA-based method~\cite{Rong2006-JFAVoronoi,Rong2011-JFACVT} processes a GPU texture by identifying the nearest site for each pixel, generating approximate Voronoi diagrams even when all pixels are correctly labeled. As mentioned above, parallel implementations of the \textit{k}NN-based method have been proposed~\cite{Ray2018-MeshlessVoronoi,Han2017-GPURVD,Liu2022-GPU3DVoronoi,Basselin2021-GPUknnRPD}. Specifically, the algorithm of Ray et al.~\cite{Ray2018-MeshlessVoronoi} works on simple domains, while those of Han et al.~\cite{Han2017-GPURVD} and Liu et al.~\cite{Liu2022-GPU3DVoronoi} are designed for surface meshes and tetrahedral meshes, respectively. To leverage the GPU architecture, these algorithms typically assign a fixed and sufficiently large number of neighbors to each site, imposing a high memory overhead on the GPU. To reduce memory usage, Basselin et al.~\cite{Basselin2021-GPUknnRPD} presented a multi-pass parallel scheme, starting with a small number of neighbors and increasing it iteratively for Voronoi cells that do not meet the security radius in subsequent passes.

Our method can be easily extended to GPUs. The procedure for generating simplex-cell intersections is adapted to allow independent computation, thereby enabling a parallel algorithm that operates without inter-thread data synchronization on GPUs. Compared to the aforementioned \textit{k}NN-based implementations, our parallel algorithm not only uses less memory but also delivers faster results.

\section{Preliminaries}
\label{sec:preliminaries}

\subsection{Clipped or restricted Voronoi diagram}

Given a set of sites $\mathcal{V} = \left \{ \mathbf{v}_i \in \mathbb{R}^d \right \}_{i = 1}^n$, the corresponding Voronoi cells are defined as 
\begin{equation}
\label{equ:voronoi-cell}
    \Omega_i = \left \{ \mathbf{x} \in \mathbb{R}^d | \|\mathbf{x}-\mathbf{v}_i\| \leq \| \mathbf{x}-\mathbf{v}_j\|, \forall j \neq i \right \}, i,\dots,n.
\end{equation}
Suppose that the given domain $\mathcal{M}$ consists of a set of small simplices, i.e., $\mathcal{M} = \left \{ \mathcal{T}_a \right \}_{a = 1}^m$, our goal is to compute the intersections $\left \{ \Omega_{i|\mathcal{M}} \right \}_{i = 1}^n$ of the Voronoi cells and the domain, i.e., each site $\mathbf{v}_i$ dominates a sub-domain $\Omega_{i|\mathcal{M}} = \Omega_i \bigcap \mathcal{M}, i = 1, \dots, n$. The set of domain-cell intersections $\left \{ \Omega_{i|\mathcal{M}} \right \}_{i = 1}^n$ provides a partition of the domain, it is called clipped Voronoi diagram if $\mathcal{M}$ is volumetric, or restricted Voronoi diagram if $\mathcal{M}$ is a surface mesh.

\subsection{Edge/face flags of simplex-cell intersection}

Our task is to compute a set of simplex-cell intersections as fast as possible. This requires an efficient search for pairs of simplices and Voronoi cells that intersect with each other. Such simplex-cell pairs can be easily found by a propagation approach~\cite{Yan2009-Remeshing, Yan2013-ClippedVoronoi}. 

\begin{figure}[t]
    \centering
    \subfloat[]{\includegraphics[height=0.15\textwidth]{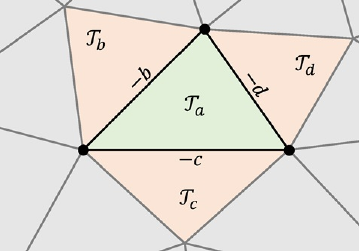}} \quad
    \subfloat[]{\includegraphics[height=0.15\textwidth]{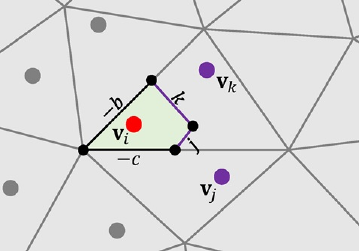}}
    \caption{Each edge/face flag indicates the adjacent domain simplex ($<0$) or the neighboring Voronoi cell ($>0$). (a) edge/face flags of a domain simplex $\mathcal{T}_a \in \mathcal{M}$; (b) edge/face flags of the intersection $\mathcal{P}_{a,i}$ (light green) of $\mathcal{T}_a$ and Voronoi cell $\Omega_i$, where the large disks are sites (red, purple, grey), and the small black disks are corner points of $\mathcal{P}_{a,i}$.}
    \label{fig:geometry}
\end{figure}

The intermediate and final results of each simplex-cell intersection are both a polygon/polytope, and we can attach a flag to each edge/face of the polygon/polytope to achieve the propagation. For an edge/face of polygon/polytope $\mathcal{P}$, if it is shared by $\mathcal{P}$ and a domain simplex $\mathcal{T}_b$, its flag is set as $(-1)\times$ the index of $\mathcal{T}_b$, i.e., $-b$, while if it is shared by $\mathcal{P}$ and a Voronoi cell $\Omega_j$, its flag is set as the index of the cell, i.e., $j$. Thus, the edge/face flags of each simplex $\mathcal{T}_a \in \mathcal{M}$ can be initialized by using the indices of the adjacent simplices, see Fig.\ref{fig:geometry}(a). If $\mathcal{T}_a$ intersects with a Voronoi cell $\Omega_i$, the intersection $\mathcal{P}_{a, i}$ can be generated after several clippings. Each valid clipping will produce a new edge/face, its flag is set as the index of the neighboring site related to the clipping, see Fig.\ref{fig:geometry}(b). Once the intersection $\mathcal{P}_{a,i}$ is finally generated, its edge/face flags indicate either the neighboring Voronoi cells of $\Omega_i$ intersected with $\mathcal{T}_a$, or the adjacent simplices of $\mathcal{T}_a$ intersected with $\Omega_i$, so that other simplex-cell pairs can be obtained quickly using edge/face flags of $\mathcal{P}_{a,i}$.

\begin{figure}[!t]
\centering
\subfloat[$\mathbf{q} \notin \Omega_i$]{\includegraphics[height=0.13\textwidth]{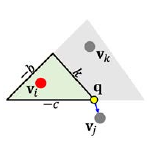}} \quad
\subfloat[$\mathbf{q} \in \Omega_i$]{\includegraphics[height=0.13\textwidth]{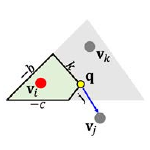}} \quad
\subfloat[$\mathbf{q} \in \Omega_i$]{\includegraphics[height=0.13\textwidth]{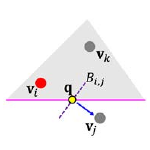}}
\caption{In our method, the point-in-cell test only needs to examine two types of points: the corner points of a polygon/polytope (a–b) and the intersecting points between a bisector and an edge (c). Each blue arrow shows the nearest site $\mathbf{v}$ of a point $\mathbf{q}$.}
\label{fig:point-in-cell}
\end{figure}

\begin{figure*}[!t]
\centering
\subfloat[]{\includegraphics[height=0.12\textwidth]{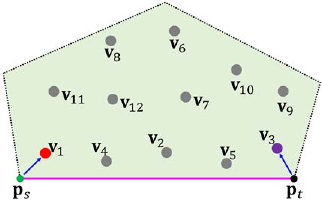}}
\subfloat[]{\includegraphics[height=0.12\textwidth]{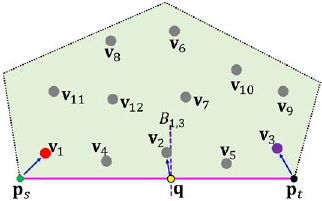}}
\subfloat[]{\includegraphics[height=0.12\textwidth]{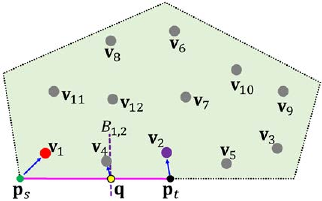}}
\subfloat[]{\includegraphics[height=0.12\textwidth]{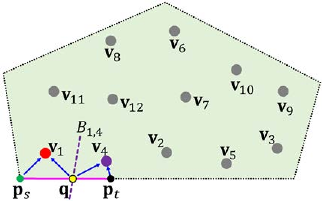}}
\subfloat[]{\includegraphics[height=0.12\textwidth]{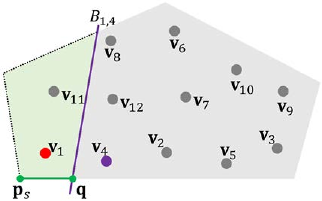}}
\caption{Illustration of the edge-based bisector search, i.e., find a target bisector on edge $\mathbf{p}_s\mathbf{p}_t$ from a polygon/polytope. Initially, $\mathbf{p}_s \in \Omega_i, \mathbf{p}_t \in \Omega_j$ (a), we iteratively compute the intersecting point $\mathbf{q}$ (yellow) of $\mathbf{p}_s\mathbf{p}_t$ (pink) and the bisector $B_{i,j}$ (purple), and update $j$ as the index of the nearest site of $\mathbf{q}$ (b-c), until $\mathbf{q} \in \Omega_i$, then the target bisector is found (d). Using the target bisector to clip the polygon/polytope, the two endpoints of the rest of the edge are both in $\Omega_i$ (e). Each blue arrow shows the nearest site $\mathbf{v}$ of a point $\mathbf{p}$. $i = 1$ in this figure.}
\label{fig:edge-based-clip}
\end{figure*}

\begin{figure*}[!t]
\centering
\subfloat[]{\includegraphics[height=0.12\textwidth]{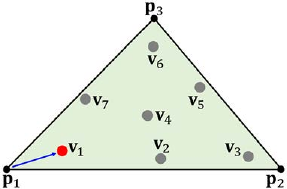}}
\subfloat[]{\includegraphics[height=0.12\textwidth]{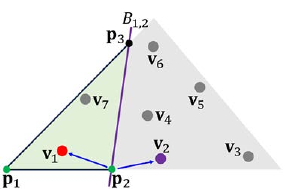}}
\subfloat[]{\includegraphics[height=0.12\textwidth]{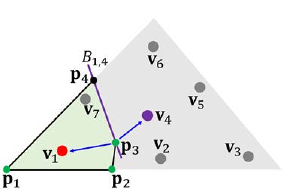}}
\subfloat[]{\includegraphics[height=0.12\textwidth]{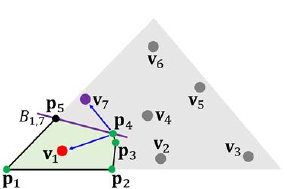}}
\subfloat[]{\includegraphics[height=0.12\textwidth]{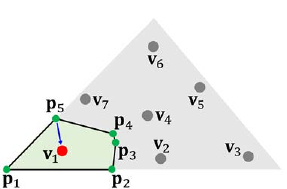}}
\caption{Illustration of computing the intersection $\mathcal{P}$ of a simplex $\mathcal{T} \in \mathcal{M}$ and a Voronoi cell $\Omega_i$. At least one corner point of $\mathcal{T}$ should be in $\Omega_i$. $\mathcal{P}$ is first initialized as $\mathcal{T}$. Select an edge whose starting point is in $\Omega_i$ but the end point is not, then find a bisector by using the proposed edge-based search scheme to clip $\mathcal{P}$. Repeat the selection and clipping steps until all the corner points are in $\Omega_i$. $i = 1$ in this figure.}
\label{fig:region-cell}
\end{figure*}

\begin{figure*}[!t]
\centering
\subfloat[]{\includegraphics[height=0.13\textwidth]{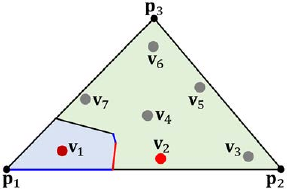}}
\subfloat[]{\includegraphics[height=0.13\textwidth]{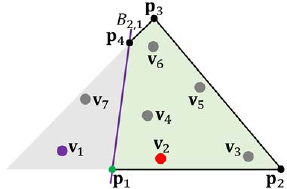}}
\subfloat[]{\includegraphics[height=0.13\textwidth]{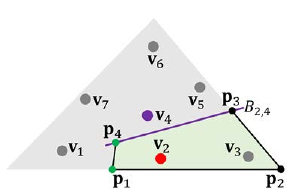}}
\subfloat[]{\includegraphics[height=0.13\textwidth]{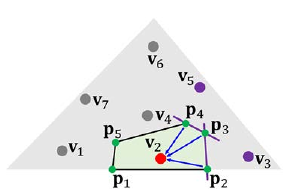}}
\caption{Illustration of computing the intersection $\mathcal{P}_{a,j}$ of a simplex $\mathcal{T}_a \in \mathcal{M}$ and a Voronoi cell $\Omega_j$, where there is no corner point of $\mathcal{T}_a$ being in $\Omega_j$. (a) The intersection $\mathcal{P}_{a,u}$ (light blue polygon) of $\mathcal{T}_a$ and $\Omega_u$ has been obtained, one of its edges/faces (red segment) involves a clipping related to the bisector $B_{u,j}$, and its adjacent edges/faces (blue segments) are useful to construct an intermediate polygon/polytope $\mathcal{G}_{a,j}$; (b-c) construction of $\mathcal{G}_{a,j}$; (d) computation of $\mathcal{P}_{a,j}$ from $\mathcal{G}_{a,j}$ by using Algorithm \ref{alg:region-cell}. $u = 1, j = 2$ in this figure.}
\label{fig:propagation}
\end{figure*}

\section{The Proposed Method}
\label{sec:proposed-method}

Benefiting from the convexity of Voronoi cells, it is known that a polygon/polytope lies entirely in a Voronoi cell if all its corner points are in the cell. If a simplex $\mathcal{T}_a \in \mathcal{M}$ intersects with a Voronoi cell $\Omega_i$, our strategy is to remove the part of $\mathcal{T}_a$ outside $\Omega_i$, ensuring all the corner points of the rest are in $\Omega_i$.

Each simplex can be decomposed into smaller polygons/polytopes, each belonging to a Voronoi cell, the details are provided in \ref{sec:serial-algorithm}. Because all Voronoi cells are unknown, how to determine whether a space point is in a target Voronoi cell is the key to our method. We set up a test called point-in-cell in \ref{sec:point-in-cell} to accomplish the determination. By using the tests, we present a novel solution that only involves valid clippings to generate a simplex-cell intersection in \ref{sec:region-cell}, and the bisector of each clipping can be found by an edge-based search scheme introduced in \ref{sec:bisector-search}.

\subsection{The point-in-cell test}
\label{sec:point-in-cell}

Given a query point $\mathbf{q}$ and the index $i$ of a Voronoi cell $\Omega_i$, the point-in-cell test tells us whether the point is in the cell or not, which is frequently invoked in our method.

Based on the definition of Voronoi cells, $\mathbf{q}$ is in $\Omega_i$ if the distance from $\mathbf{q}$ to $\mathbf{v}_i$ is not greater than that from $\mathbf{q}$ to any other sites. Let $\mathcal{N}_\mathbf{q}$ be the nearest site of $\mathbf{q}$. We can conclude that $\mathbf{q} \in \Omega_i$ if $\mathcal{N}_\mathbf{q} = \mathbf{v}_i$ or if the following equation holds:
\begin{equation}
    \label{equ:point-in-cell-condition}
    \| \mathbf{q} - \mathbf{v}_i \| = \| \mathbf{q} - \mathcal{N}_\mathbf{q} \|.
\end{equation}
Thus, only one time of nearest neighbor search is required in each point-in-cell test. 

However, due to the finite precision of floating-point arithmetic in computers, directly applying Equation (\ref{equ:point-in-cell-condition}) is unreliable when $\mathcal{N}_\mathbf{q} \neq \mathbf{v}_i$. To ensure robustness, we therefore perform this judgment using integer flags around $\mathbf{q}$. In our solution for computing the simplex-cell intersection $\mathcal{P}$ of a domain simplex $\mathcal{T}$ and $\Omega_i$ (Section \ref{sec:region-cell}), the point-in-cell test only needs to examine two types of points: (1) the corner points of $\mathcal{P}$ (see Fig.~\ref{fig:point-in-cell}(a-b)); (2) the intersecting points between a bisector and an edge of $\mathcal{P}$ (see Fig.~\ref{fig:point-in-cell}(c)).

For a corner point $\mathbf{q}$ of $\mathcal{P}$, we collect all flags $\mathcal{F}_\mathbf{q}$ of edges/faces incident to $\mathbf{q}$, and if the index of $\mathcal{N}_\mathbf{q}$ appears in $\mathcal{F}_\mathbf{q}$, then $\mathbf{q} \in \Omega_i$ for the case $\mathcal{N}_\mathbf{q} \neq \mathbf{v}_i$. For example in Fig.~\ref{fig:point-in-cell}(a), $\mathcal{F}_\mathbf{q} = \left \{ -c, k \right \}$, $\mathcal{N}_\mathbf{q} = \mathbf{v}_j$, because $j \notin \mathcal{F}_\mathbf{q}$, then $\mathbf{q} \notin \Omega_i$; while in Fig.~\ref{fig:point-in-cell}(b), $\mathcal{F}_\mathbf{q} = \left \{ j, k \right \}$, $\mathcal{N}_\mathbf{q} = \mathbf{v}_j$, we have $\mathbf{q} \in \Omega_i$ since $j \in \mathcal{F}_\mathbf{q}$.

For an intersecting point $\mathbf{q}$ between a bisector $B_{i,j}$ of sites $\mathbf{v}_i$, $\mathbf{v}_j$ and an edge, if $\mathcal{N}_{\mathbf{q}} = \mathbf{v}_j$ (Fig.~\ref{fig:point-in-cell}(c)), it is known that $\mathbf{q}$ must be on the common edge of Voronoi cells $\Omega_i$ and $\Omega_j$, namely, $\mathbf{q} \in  \Omega_i$.


\begin{algorithm}[t]
\caption{Computation of simplex-cell intersection}
\label{alg:region-cell}
\begin{algorithmic}[1]
\REQUIRE a simplex $\mathcal{T} \in \mathcal{M}$ and a site index $i$.
\ENSURE intersection $\mathcal{P}$ of $\mathcal{T}$ and $\Omega_i$.
\STATE $\mathcal{P} \leftarrow \mathcal{T}$;
\WHILE {$\mathcal{P}$ has corner point being outside $\Omega_i$}
\STATE select an edge $\mathbf{p}_s\mathbf{p}_t$, requiring that $\mathbf{p}_s \in \Omega_i, \mathbf{p}_t \notin \Omega_i$;
\STATE find a bisector $B_{i,j}$ on $\mathbf{p}_s\mathbf{p}_t$ (Section \ref{sec:bisector-search});
\STATE $\mathcal{P} \leftarrow$ the result after clipping $\mathcal{P}$ using $B_{i,j}$;
\ENDWHILE
\end{algorithmic}
\end{algorithm}

\subsection{Edge-based bisector search}
\label{sec:bisector-search}

Consider an edge $\mathbf{p}_s\mathbf{p}_t$ from a polygon/polytope. Suppose that the starting point $\mathbf{p}_s$ is in the Voronoi cell $\Omega_i$ and the end point $\mathbf{p}_t$ is in the cell $\Omega_j$. We aim to find a target bisector, after which is used to clip the polygon/polytope, the two endpoints of the rest of the edge are both in $\Omega_i$. If $j = i$, the whole edge is in $\Omega_i$ and no clipping is needed, here we focus on the case that $j \neq i$.

It is known that the points shared by two neighboring Voronoi cells $\Omega_u, \Omega_w$ must be on the bisector $B_{u,w}$ of the two corresponding sites $\mathbf{v}_u, \mathbf{v}_w$, and the clipping using $B_{u,w}$ must contribute to the computation of Voronoi cells $\Omega_u, \Omega_w$. This observation inspires us the way to search for the target bisector on edge $\mathbf{p}_s\mathbf{p}_t$.

Initially, $\mathbf{p}_s \in \Omega_i, \mathbf{p}_t \in \Omega_j$, we construct the bisector $B_{i,j}$ of the corresponding sites $\mathbf{v}_i$ and $\mathbf{v}_j$, and compute the intersecting point $\mathbf{q}$ of $\mathbf{p}_s\mathbf{p}_t$ and $B_{i,j}$. By using the point-in-cell test (section \ref{sec:point-in-cell}), we can easily obtain the relationship between $\mathbf{q}$ and $\Omega_i$: (1) If the test answers false, and the nearest site of $\mathbf{q}$ is denoted as $\mathbf{v}_k$, then $k$ must be different with $i$ and $j$, which means that the clipping using $B_{i,j}$ must be useless. In this case, we update $j \leftarrow k, \mathbf{p}_t \leftarrow \mathbf{q}$ and repeat the above steps, i.e., update $B_{i,j}$ and $\mathbf{q}$, and redo the point-in-cell test. (2) If the test answers true, we know that the whole segment $\mathbf{p}_s\mathbf{q}$ is in the cell $\Omega_i$, then the newest $B_{i,j}$ is the target bisector, which can be directly used to clip the polygon/polytope. Fig. \ref{fig:edge-based-clip} shows an illustration.

\begin{algorithm}[t]
\caption{Computation of clipped or restricted VD}
\label{alg:serial}
\begin{algorithmic}[1]
\REQUIRE sites $\mathcal{V} = \left \{ \mathbf{v}_i \right \}_{i = 1}^n$ and domain $\mathcal{M} = \left \{ \mathcal{T}_a \right \}_{a = 1}^m$
\ENSURE clipped or restricted Voronoi cells $\left \{ \Omega_{i|\mathcal{M}} \right \}_{i = 1}^n$.
\STATE build a structure $\mathcal{S}$ of $\mathcal{V}$ for nearest neighbor searches (section~\ref{sec:NNS-grids});
\STATE $\Omega_{i|\mathcal{M}} \leftarrow \emptyset, i=1,\dots,n$;
\FOR {each $\mathcal{T}_a \in \mathcal{M}$}
\STATE $\mathbf{v}_i \leftarrow$ the nearest site of the first corner point of $\mathcal{T}_a$;
\STATE compute intersection $\mathcal{P}_{a,i}$ of $\mathcal{T}_a$ and $\Omega_i$ (Alg. \ref{alg:region-cell});
\STATE $\Omega_{i|\mathcal{M}} \leftarrow \Omega_{i|\mathcal{M}} \bigcup \mathcal{P}_{a,i}$;
\STATE $Q \leftarrow \emptyset$;
\STATE add other sites to $Q$ using edge/face flags of $\mathcal{P}_{a,i}$;
\WHILE {$Q$ is not empty}
\STATE $\mathbf{v}_j \leftarrow$ popped from $Q$;
\STATE construct a polygon/polytope $\mathcal{G}_{a,j}$ from $\mathcal{T}_a$;
\STATE compute intersection $\mathcal{P}_{a,j}$ of $\mathcal{G}_{a,j}$ and $\Omega_j$ (Alg. \ref{alg:region-cell});
\STATE $\Omega_{j|\mathcal{M}} \leftarrow \Omega_{j|\mathcal{M}} \bigcup \mathcal{P}_{a,j}$;
\STATE add unvisited sites to $Q$ using edge/face flags of $\mathcal{P}_{a,j}$;
\ENDWHILE
\ENDFOR
\end{algorithmic}
\end{algorithm}

The target bisector on the edge can be found after several point-in-cell tests. The number of tests is proportional to the number of Voronoi cells intersecting the edge. To reduce the number of tests, a minor modification can be made before searching for the target bisector, which will be discussed later in Section \ref{sec:new-bisector-search}.

\subsection{Computation of simplex-cell intersection}
\label{sec:region-cell}

Suppose that at least one corner point of a simplex $\mathcal{T}$ is in the Voronoi cell $\Omega_i$ of site $\mathbf{v}_i$, we present a novel solution for generating the intersection $\mathcal{P}$ of $\mathcal{T}$ and $\Omega_i$. Specifically, $\mathcal{P}$ is initialized as $\mathcal{T}$, and it is iteratively removed the part outside $\Omega_i$. The steps are as follows: (1) select an edge $\mathbf{p}_s\mathbf{p}_t$ from $\mathcal{P}$, requiring that $\mathbf{p}_s \in \Omega_i, \mathbf{p}_t \notin \Omega_i$; (2) find a bisector $B_{i,j}$ by using the proposed edge-based search scheme (Section \ref{sec:bisector-search}) on $\mathbf{p}_s\mathbf{p}_t$, and use it to clip $\mathcal{P}$, the result is still denoted as $\mathcal{P}$. Repeat the selection and clipping steps until $\mathcal{P}$ has no edge meeting the requirement in step 1, which implies that all the corner points of $\mathcal{P}$ are in $\Omega_i$, i.e., the simplex-cell intersection is finally obtained. The steps are also listed in Algorithm \ref{alg:region-cell}, and an illustration is given in Fig. \ref{fig:region-cell}. 

At most one clipping is involved for each edge of the intersection, and all the clippings are contributed to the final result. This is a great advantage of our solution, since a part of the clippings in both DT-based and \textit{k}NN-based methods could be useless for generating such intersections. Furthermore, when clipping $\mathcal{P}$ with a bisector, a part of corner points of $\mathcal{P}$ are guaranteed to lie in the Voronoi cell $\Omega_i$. Thus, it is unnecessary to check whether these corner points belong to the half-space containing $\mathbf{v}_i$, which reduces computational effort.

\subsection{Computation of clipped or restricted Voronoi diagram}
\label{sec:serial-algorithm}

We now concentrate on the computation of clipped or restricted Voronoi cells $\left \{ \Omega_{i|\mathcal{M}} \right \}_{i=1}^n$. Initially, they are set to empty, i.e., $\Omega_{i|\mathcal{M}}=\emptyset, i = 1, \dots, n$. We can decompose each domain simplex into smaller polygons/polytopes, each of which is entirely in a Voronoi cell such that it can be added to the corresponding clipped or restricted cell. Since the decomposition process is identical for all simplices, we take $\mathcal{T}_a \in \mathcal{M}$ as an example to describe our method. The pseudo-code can be found in Algorithm \ref{alg:serial}.

\begin{figure*}[!t]
\centering
\subfloat[]{\includegraphics[height=0.12\textwidth]{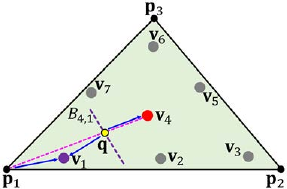}}
\subfloat[]{\includegraphics[height=0.12\textwidth]{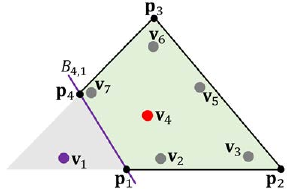}}
\subfloat[]{\includegraphics[height=0.12\textwidth]{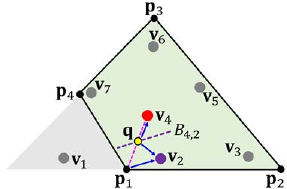}}
\subfloat[]{\includegraphics[height=0.12\textwidth]{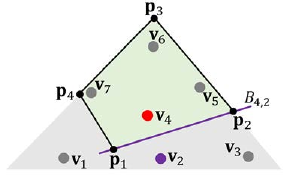}}
\subfloat[]{\includegraphics[height=0.12\textwidth]{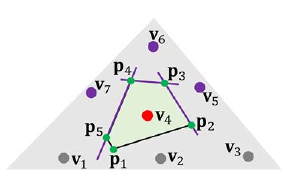}}
\caption{A modified solution for computing intersection $\mathcal{P}$ of a simplex $\mathcal{T}$ and a Voronoi cell $\Omega_i$. After $\mathcal{P}$ is initialized as $\mathcal{T}$, it is clipped iteratively. In each iteration, we select a corner point of $\mathcal{P}$ that is outside $\Omega_i$ and build a virtual edge (pink dotted line) by connecting $\mathbf{v}_i$ and the corner point, based on which we can find a bisector by using the proposed edge-based search scheme to clip $\mathcal{P}$. After a few iterations, the final $\mathcal{P}$ has no corner point being outside $\Omega_i$. The computation is independent for each site and suitable for parallel computing. $i = 4$ in this figure.}
\label{fig:parallel-clips}
\end{figure*}

\subsubsection{First polygon/polytope}
\label{sec:first-decomposition}

At the beginning, we have to find the nearest site $\mathbf{v}_i$ of the first corner point of $\mathcal{T}_a$, indicating that $\mathcal{T}_a$ must intersect with the Voronoi cell $\Omega_i$. The intersection $\mathcal{P}_{a,i}$ of $\mathcal{T}_a$ and $\Omega_i$ can be directly generated by using Algorithm \ref{alg:region-cell}, and it is then added to $\Omega_{i|\mathcal{M}}$, i.e, $\Omega_{i|\mathcal{M}} = \Omega_{i|\mathcal{M}} \bigcup \mathcal{P}_{a,i}$.

\subsubsection{Propagation}

It has to further compute the intersections of $\mathcal{T}_a$ and other Voronoi cells, the simplex-cell pairs can be obtained by using the edge/face flags of $\mathcal{P}_{a,i}$. Specifically, we set up a queue $Q$ to store the unvisited sites whose Voronoi cells are intersected with $\mathcal{T}_a$. After the first polygon/polytope $\mathcal{P}_{a,i}$ is generated, we traverse its all edges/faces. For each edge/face, if it has a positive flag $k$, meaning that $\Omega_k$ must intersect with $\mathcal{T}_a$, we add $\mathbf{v}_k$ to the queue. Iteratively, a site $\mathbf{v}_j$ is popped from $Q$, and we compute the intersection $\mathcal{P}_{a,j}$ of $\mathcal{T}_a$ and $\Omega_j$, which is then added to $\Omega_{j|\mathcal{M}}$. The queue $Q$ can be further updated by adding more unvisited sites using the edge/face flags of $\mathcal{P}_{a,j}$. The procedure is terminated once the queue is empty, and the decomposition on $\mathcal{T}_a$ is completed.

The last obstacle is the computation of $\mathcal{P}_{a,j}$, where $j$ is the index of the site popped from the queue in each iteration. If there is at least one corner point of $\mathcal{T}_a$ being in the Voronoi cell $\Omega_j$, Algorithm \ref{alg:region-cell} can be directly applied to compute $\mathcal{P}_{a,j}$. Otherwise, no edge can be selected from $\mathcal{T}_a$ to launch clippings. Therefore, an intermediate polygon/polytope $\mathcal{G}_{a,j}$ that has at least one corner point being in $\Omega_j$ should be constructed before computing $\mathcal{P}_{a,j}$.

The steps for constructing $\mathcal{G}_{a, j}$ are as follows. Suppose that the intersection $\mathcal{P}_{a,u}$ of $\mathcal{T}_a$ and $\Omega_u$ has been obtained, and its edge/face $f$ has a positive flag $j$, leading to that $\mathbf{v}_j$ is added to the queue $Q$. The tuple $(u, f)$ can be used to construct $\mathcal{G}_{a,j}$ once $\mathbf{v}_j$ is popped from the queue. Initially, copy $\mathcal{T}_a$ to $\mathcal{G}_{a,j}$. For the edge/face $f$ of $\mathcal{P}_{a,u}$, because it relates to the bisector $B_{u,j}$, we can directly use $B_{j,u}$ to clip $\mathcal{G}_{a,j}$. Furthermore, we have to check each adjacent edge/face of $f$ in $\mathcal{P}_{a,u}$, if it has a positive flag $w$, then $B_{j,w}$ can be used to clip $\mathcal{G}_{a,j}$. The result after each clipping on $\mathcal{G}_{a,j}$ is still denoted as $\mathcal{G}_{a,j}$. After these clippings, the final $\mathcal{G}_{a,j}$ is obtained, an illustration is given in Fig. \ref{fig:propagation}(a-c). Next, $\mathcal{G}_{a,j}$ can be fed to Algorithm \ref{alg:region-cell} to compute $\mathcal{P}_{a,j}$, shown in Fig. \ref{fig:propagation}(d).

\begin{algorithm}[t]
\caption{Modified computation of simplex-cell intersection}
\label{alg:independent-region-cell}
\begin{algorithmic}[1]
\REQUIRE a simplex $\mathcal{T}$ and the site index $i$.
\ENSURE intersection $\mathcal{P}$ of $\mathcal{T}$ and $\Omega_i$.
\STATE $\mathcal{P} \leftarrow \mathcal{T}$;
\WHILE {$\mathcal{P}$ has corner point $\mathbf{p}$ being outside $\Omega_i$}
\STATE find a bisector $B_{i,j}$ on virtual edge $\mathbf{v}_i\mathbf{p}$ (Section \ref{sec:bisector-search});
\STATE $\mathcal{P} \leftarrow$ the result after clipping $\mathcal{P}$ using $B_{i,j}$;
\ENDWHILE
\end{algorithmic}
\end{algorithm}

\section{Parallel Computation on GPUs}
\label{sec:parallel-computation}

In this section, we extend our method to GPUs for the parallel computation of clipped or restricted Voronoi cells. To achieve this, an independent solution for generating each simplex-cell intersection is required. The solution presented in Section \ref{sec:region-cell} becomes inapplicable here. We therefore propose a modified version, which will serve as the foundation for the parallel algorithms introduced subsequently.

\subsection{Modified computation of simplex-cell intersection}

Because each site must be in its own Voronoi cell, to generate the intersection $\mathcal{P}$ of a simplex $\mathcal{T}$ and a Voronoi cell $\Omega_i$, we can build virtual edges by connecting the corresponding site $\mathbf{v}_i$ and the corner points of $\mathcal{T}$. These virtual edges can then be fed into the proposed edge-based search scheme to find bisectors. Specifically, $\mathcal{P}$ is first initialized as $\mathcal{T}$, and it is clipped iteratively to be smaller and smaller. In each iteration, two steps are executed: (1) select a corner point $\mathbf{p}$ of $\mathcal{P}$, requiring that $\mathbf{p} \notin \Omega_i$; (2) find a bisector of $\mathbf{v}_i$ and other site by using the edge-based search scheme on virtual edge $\mathbf{v}_i\mathbf{p}$, and use it to clip $\mathcal{P}$, the result is still denoted as $\mathcal{P}$. After a few iterations, the final $\mathcal{P}$ has no corner point being outside $\Omega_i$. Fig. \ref{fig:parallel-clips} shows an illustration, and Algorithm \ref{alg:independent-region-cell} lists the pseudo-code. This modified algorithm is independent of each site and suitable for parallel computing.

It is worth pointing out that applying the modified computation of simplex-cell intersection in serial computing is not suggested, since it will result in efficiency loss. For generating the intersection $\mathcal{P}$ of $\mathcal{T}$ and $\Omega_i$, the original solution presented in \ref{sec:region-cell} ensures that at least one of the new corner points introduced by each clipping must be in $\Omega_i$, which is not supported in the modified solution. Consequently, the number of the point-in-cell tests in the modified solution is more than that in the original solution.

\subsection{Parallel computation on simple domains}
\label{sec:parallel-regular}

If the given domain is simple with only a few corner points (e.g., a cube in $\mathbb{R}^3$), the parallel computation of clipped or restricted Voronoi cells is straightforward. We assign one thread for each site, the main task of each thread is to iteratively clip the domain by using Algorithm \ref{alg:independent-region-cell}.

Our parallel algorithm guarantees exact results because Algorithm \ref{alg:independent-region-cell} employs an iterative process that continues until all corner points are in the target Voronoi cell. In addition, similar to the \textit{k}NN-based method, our parallel algorithm requires no inter-thread data synchronization on GPUs, enabling high parallelism in computation.

\subsection{Parallel computation on complex domains}

If the given domain is complex and consists of a set of small simplices, the parallel computation of clipped or restricted Voronoi diagrams becomes much more complicated. Here, we focus on the surface and 3D cases.

\subsubsection{Surface case}
\label{sec:parallel-rvd}

Suppose that the given domain is a surface mesh composed of $m$ triangles. To compute restricted Voronoi cells in parallel, our strategy identifies triangle-cell pairs, each assigned to a single thread. The primary task of each thread is therefore to compute triangle-cell intersection using Algorithm~\ref{alg:independent-region-cell}. 

Specifically, for each triangle $\mathcal{T}_a \in \mathcal{M}$, we identify the nearest site $\mathbf{v}_i$ of its centroid, forming a pair $(\mathcal{T}_a, \mathbf{v}_i)$. These pairs can then be assigned to GPU threads for computing corresponding triangle-cell intersections. Based on positive edge flags of the intersections, additional triangle-cell pairs can be discovered, similar to the propagation scheme of Han et al.~\cite{Han2017-GPURVD}. The computation terminates when no more triangle-cell pairs are obtained. Algorithm~\ref{alg:parallel-RVD} outlines the steps.

\begin{algorithm}[t]
\caption{Parallel computation of restricted VD}
\label{alg:parallel-RVD}
\begin{algorithmic}[1]
\REQUIRE sites $\mathcal{V} = \left \{ \mathbf{v}_i \right \}_{i = 1}^n$, domain triangles $\mathcal{M} = \left \{ \mathcal{T}_a \right \}_{a = 1}^m$.
\ENSURE restricted Voronoi cells $\left \{ \Omega_{i|\mathcal{M}} \right \}_{i = 1}^n$.
\STATE build a structure $\mathcal{S}$ of $\mathcal{V}$ for nearest neighbor searches (section~\ref{sec:NNS-grids});
\STATE initialize two lists $\mathcal{L}_1 \leftarrow \emptyset, \mathcal{L}_2 \leftarrow \emptyset$;
\FOR{each triangle $\mathcal{T}_a \in \mathcal{M}$ in parallel}
\STATE $\mathbf{v}_i \leftarrow$ the nearest site of the centroid of $\mathcal{T}_a$;
\STATE insert $(\mathcal{T}_a, \mathbf{v}_i)$ to $\mathcal{L}_1$ and label it in atomic;
\ENDFOR
\WHILE {$\mathcal{L}_1$ is not empty}
\STATE $\mathcal{L}_2 \leftarrow \emptyset$;
\FOR {each pair $(\mathcal{T}_a, \mathbf{v}_i) \in \mathcal{L}_1$ in parallel}
\STATE compute intersection $\mathcal{P}_{a,i}$ of $\mathcal{T}_a$ and $\Omega_i$ (Alg. \ref{alg:independent-region-cell});
\STATE $\Omega_{i|\mathcal{M}} \leftarrow \Omega_{i|\mathcal{M}} \bigcup \mathcal{P}_{a,i}$ in atomic;
\FOR{each edge $f$ of $\mathcal{P}_{a,i}$}
\IF{$f$'s flag $j > 0$ and $(\mathcal{T}_a, \mathbf{v}_j)$ is unlabeled}
\STATE insert $(\mathcal{T}_a, \mathbf{v}_j)$ to $\mathcal{L}_2$ and label it in atomic;
\ENDIF
\ENDFOR
\ENDFOR
\STATE $\mathcal{L}_1 \leftarrow \mathcal{L}_2$;
\ENDWHILE
\end{algorithmic}
\end{algorithm}

\subsubsection{3D case}

The domain is composed of $m$ tetrahedra in this case. Because we expect exact results, the propagation scheme used in steps 9$\sim$17 in Algorithm~\ref{alg:parallel-RVD} is also applied to the parallel computation of 3D clipped Voronoi cells. In fact, only a subset of the sites needs to be processed in this manner, while the others can be processed as in Section \ref{sec:parallel-regular}.

By computing restricted Voronoi cells, the given sites $\mathcal{V}$ can be divided into two categories~\cite{Yan2013-ClippedVoronoi}: (1) boundary sites $\mathcal{V}_\mathcal{B}$ with non-empty restricted Voronoi cells, and (2) interior sites $\mathcal{V}_\mathcal{I} = \mathcal{V} \setminus \mathcal{V}_\mathcal{B}$. For an interior site $\mathbf{v}_i \in \mathcal{V}_\mathcal{I}$, its domain-cell intersection is identical to the Voronoi cell itself, i.e., $\Omega_{i|\mathcal{M}} = \Omega_i$. Thus, the parallel algorithm in Section \ref{sec:parallel-regular} can be directly applied to interior sites.

For boundary sites, the parallel computation of clipped Voronoi cells are as follows. For each site $\mathbf{v}_j \in \mathcal{V}_{\mathcal{B}}$, its restricted Voronoi cell consists of multiple polygons, each originating from a boundary triangle $\Delta$ of the domain. Thus, a pair $(\mathcal{T}_a, \mathbf{v}_j)$ is formed, where $\mathcal{T}_a$ represents the incident tetrahedron of $\Delta$. After initializing a set of tetrahedron-cell pairs, the remaining steps follow those described in Section \ref{sec:parallel-rvd}, but negative face flags of the generated tetrahedron-cell intersections are used to propagate tetrahedron-cell pairs. Algorithm~\ref{alg:parallel-ClippedVD} provides the detailed procedure.

\begin{algorithm}[t]
\caption{Parallel computation of 3D clipped VD}
\label{alg:parallel-ClippedVD}
\begin{algorithmic}[1]
\REQUIRE sites $\mathcal{V} = \left \{ \mathbf{v}_i \right \}_{i = 1}^n$, domain tetrahedra $\mathcal{M} = \left \{ \mathcal{T}_a \right \}_{a = 1}^m$.
\ENSURE 3D clipped Voronoi cells $\left \{ \Omega_{i|\mathcal{M}} \right \}_{i = 1}^n$.
\STATE build a structure $\mathcal{S}$ of $\mathcal{V}$ for nearest neighbor searches (section~\ref{sec:NNS-grids});\\
// \texttt{steps 2 $\sim$ 5 are for dividing sites}
\STATE $\mathcal{M}_\mathcal{B} \leftarrow$ boundary triangles of the domain;
\STATE compute restricted Voronoi cells $\left \{ \mathcal{C}_{i} \right \}_{i = 1}^n$ on $\mathcal{M}_\mathcal{B}$ (Alg. \ref{alg:parallel-RVD});
\STATE boundary sites $\mathcal{V}_\mathcal{B} \leftarrow \left \{ \mathbf{v}_i \lvert \mathcal{C}_i \neq \emptyset \right \} $;
\STATE interior sites $\mathcal{V}_\mathcal{I} \leftarrow \mathcal{V} \setminus \mathcal{V}_\mathcal{B}$;\\
// \texttt{cells of interior sites}
\STATE given $\mathcal{M}$'s bounding box, compute cells of interior sites (section~\ref{sec:parallel-regular});\\
// \texttt{cells of boundary sites}
\STATE initialize two lists $\mathcal{L}_1 \leftarrow \emptyset, \mathcal{L}_2 \leftarrow \emptyset$;
\FOR{each site $\mathbf{v}_j \in \mathcal{V}_\mathcal{B}$ in parallel}
\FOR{each polygon $\mathcal{H} \in \mathcal{C}_j$}
\STATE $\Delta \leftarrow$ the boundary triangle that $\mathcal{H}$ originates from;
\STATE $\mathcal{T}_a \leftarrow$ the incident tetrahedron of $\Delta$;
\STATE insert $(\mathcal{T}_a, \mathbf{v}_j)$ to $\mathcal{L}_1$ and label it in atomic;
\ENDFOR
\ENDFOR
\WHILE {$\mathcal{L}_1$ is not empty}
\STATE $\mathcal{L}_2 \leftarrow \emptyset$;
\FOR {each pair $(\mathcal{T}_a, \mathbf{v}_j) \in \mathcal{L}_1$ in parallel}
\STATE compute intersection $\mathcal{P}_{a,j}$ of $\mathcal{T}_a$ and $\Omega_j$ (Alg. \ref{alg:independent-region-cell});
\STATE $\Omega_{j|\mathcal{M}} \leftarrow \Omega_{j|\mathcal{M}} \bigcup \mathcal{P}_{a,j}$ in atomic;
\FOR{each face $f$ of $\mathcal{P}_{a,j}$}
\IF{$f$'s flag $(-b) < 0$ and $(\mathcal{T}_{b}, \mathbf{v}_j)$ is unlabeled}
\STATE insert $(\mathcal{T}_{b}, \mathbf{v}_j)$ to $\mathcal{L}_2$ and label it in atomic;
\ENDIF
\ENDFOR
\ENDFOR
\STATE $\mathcal{L}_1 \leftarrow \mathcal{L}_2$;
\ENDWHILE
\end{algorithmic}
\end{algorithm}

\section{Implementation Details}
\label{sec:implementation-details}


\subsection{Improved edge-based bisector search}
\label{sec:new-bisector-search}

Given an edge $\mathbf{p}_s\mathbf{p}_t, \mathbf{p}_s \in \Omega_i$, the original version of edge-based bisector search presented in \ref{sec:bisector-search} initiates the search from the end point $\mathbf{p}_t$. A larger number of Voronoi cells intersected with the edge, more point-in-cell tests are required. We aim to reduce the number of tests to further enhance the efficiency of our method.

In fact, an excess of tests are needed because the end point $\mathbf{p}_t$ is far away from the Voronoi cell $\Omega_i$. Therefore, our strategy is to find a new point $\mathbf{g}$ on $\mathbf{p}_s\mathbf{p}_t$, requiring that $\mathbf{g}$ is very close to but not in $\Omega_i$, so that a much shorter segment $\mathbf{p}_s\mathbf{g}$ can be fed to the search scheme.

Let $\mathcal{N}_i$ be the nearest neighbor of $\mathbf{v}_i$ among all the sites. The distance $\delta = \left \| \mathcal{N}_i - \mathbf{v}_i \right \| / 2$ implies an approximate size of $\Omega_i$, which can be used to guide the search of point $\mathbf{g}$. Specifically, initiate $\lambda = 2, \mathbf{g} = \mathbf{p}_s + \lambda \delta \frac{\mathbf{p}_t - \mathbf{p}_s}{\left \| \mathbf{p}_t - \mathbf{p}_s \right \|}$, if the nearest site of $\mathbf{g}$ is $\mathbf{v}_i$, we increment $\lambda$ by 1 and repeat the search with  the updated $\mathbf{g}$; otherwise, the search terminates.

\begin{figure}[!t]
\centering
\subfloat[2D]{\includegraphics[height=0.15\textwidth]{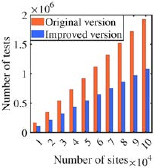}}
\subfloat[Surface]{\includegraphics[height=0.15\textwidth]{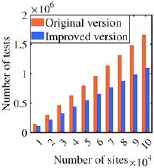}}
\subfloat[3D]{\includegraphics[height=0.15\textwidth]{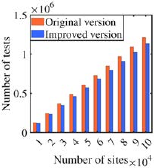}}
\caption{The numbers of point-in-cell tests invoked in Algorithm~\ref{alg:serial} using different versions of the edge-based bisector search.} 
\label{fig:reduce-tests}
\end{figure}

This simple trick reduces remarkably the number of point-in-cell tests. Fig.~\ref{fig:reduce-tests} compares the numbers of tests invoked in Algorithm~\ref{alg:serial} using original and improved versions of the edge-based bisector search. The domains are a rectangle, a surficial and a volumetric cube, respectively, and the sites are randomly sampled with varying quantities. The results clearly show that when the improved version is applied, the tests are invoked less frequently.

\subsection{Nearest neighbor search}
\label{sec:NNS-grids}

Our method heavily relies on the point-in-cell tests, each of which requires a nearest neighbor search. The KD-tree is a well-known data structure that provides efficient nearest neighbor search, which has been widely applied in many fields. However, its search efficiency remains insufficient for our method due to the massive number of required nearest neighbor searches. For example, generating 1 million Voronoi cells bounded by a cube required over 20.54 millions searches, where the sites are randomly sampled. Therefore, improving search efficiency is crucial to our method.

Because each point-in-cell test only requires the nearest site of query point, a grids-based structure~\cite{Ray2018-MeshlessVoronoi} is well-suited for this task. The given sites are first placed into a group of grids according to their coordinates. To find the nearest site for a query point, we examine nearby grids around the query point and traverse the sites within these grids to determine the closest one. More details can be found in \cite{Ray2018-MeshlessVoronoi}. We set the grid resolution so that each grid cell contains, on average, 2 $\sim$ 3 sites (for the 2D and surface cases) and 3 $\sim$ 4 sites (for the 3D case), respectively.

\begin{figure}[!t]
\centering
\subfloat[2D]{\includegraphics[height=0.15\textwidth]{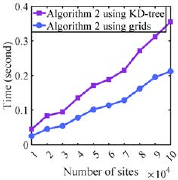}}
\subfloat[Surface]{\includegraphics[height=0.15\textwidth]{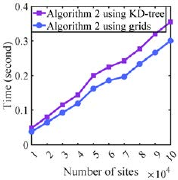}}
\subfloat[3D]{\includegraphics[height=0.15\textwidth]{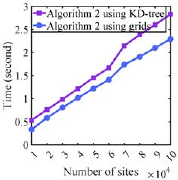}}
\caption{The runtime curves of Algorithm~\ref{alg:serial} using different structures for nearest neighbor searches.}
\label{fig:NNS-structure}
\end{figure}

By using different structures for nearest neighbor searches, Algorithm~\ref{alg:serial} demonstrates varying levels of efficiency. The runtime curves of Algorithm~\ref{alg:serial} using KD-tree and grids, respectively, are shown in Fig.~\ref{fig:NNS-structure}. The test domains include a rectangle, a surface sphere and a tetrahedralized ball, with sites randomly sampled at varying quantities. These curves indicate that Algorithm~\ref{alg:serial} achieves higher efficiency when using grids-based structure. Thus, we adopt the grids-based structure in both our serial and parallel algorithms.

\subsection{3D cell clipping on GPUs}

To accommodate the GPU architecture, we employ the 3D cell data structure proposed by Ray et al.~\cite{Ray2018-MeshlessVoronoi} when implementing Algorithm~\ref{alg:parallel-ClippedVD}. This structure comprises an array of planes and an array of corner points, where each corner point is defined as the intersection of three planes. It can also be interpreted as a triangular mesh, i.e., a plane corresponds to a mesh vertex, and a corner point corresponds to a mesh triangle. Therefore, when clipping the simplex-cell intersection $\mathcal{P}_{a,i}$ of domain simplex $\mathcal{T}_a$ and Voronoi cell $\Omega_i$ using a bisector $B_{i,j}$, the procedure consists of the following steps: (1) identify all mesh triangles $R$ of $\mathcal{P}_{a,i}$ that lie outside $\Omega_i$; (2) determine the boundary loop of $R$; (3) remove $R$ and generate new mesh triangles by connecting $B_{i,j}$ to the boundary edges of $R$. For further details, refer to \cite{Ray2018-MeshlessVoronoi}.

However, due to the finite precision of floating-point arithmetic, this cell-clipping procedure may misclassify the side of some mesh triangles and consequently generate a disconnected boundary for the set $R$. This issue cannot be avoided in the existing \textit{k}NN-based parallel algorithms~\cite{Ray2018-MeshlessVoronoi, Liu2022-GPU3DVoronoi, Basselin2021-GPUknnRPD}. In contrast, Algorithm~\ref{alg:independent-region-cell} readily avoids the problem. Because we always select a corner point $\mathbf{p}$ that is guaranteed to lie outside $\Omega_i$ for each clipping (see Fig.~\ref{fig:parallel-clips}(a, c)), we simply replace $R$ with $R_{\mathbf{p}}$ in the steps above, where $R_{\mathbf{p}}$ denotes the connected component of $R$ that contains $\mathbf{p}$.

\begin{figure*}[!t]
\centering
\subfloat[]{\includegraphics[height=0.185\textwidth]{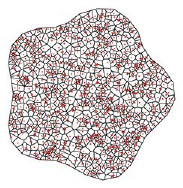}} \quad
\subfloat[]{\includegraphics[height=0.185\textwidth]{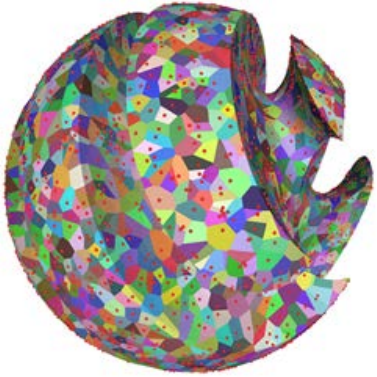}} \quad
\subfloat[]{\includegraphics[height=0.185\textwidth]{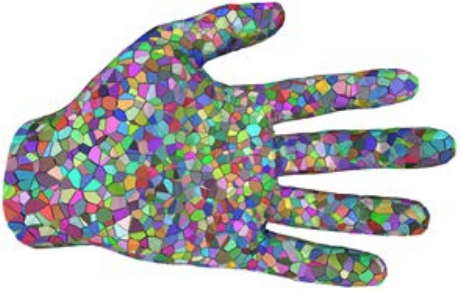}}\\
\subfloat[]{\includegraphics[height=0.185\textwidth]{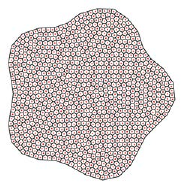}} \quad
\subfloat[]{\includegraphics[height=0.185\textwidth]{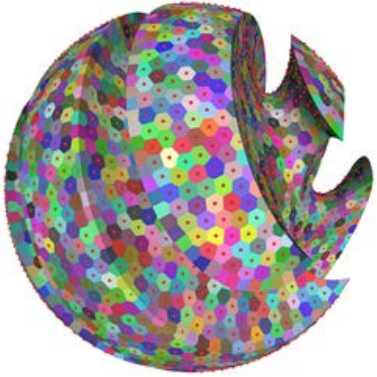}} \quad
\subfloat[]{\includegraphics[height=0.185\textwidth]{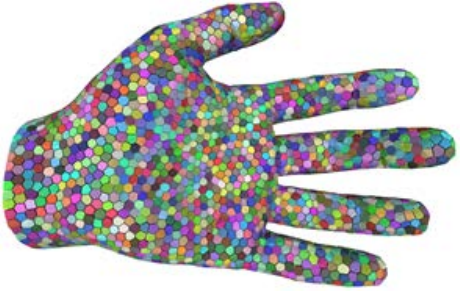}}\\
\subfloat[]{\includegraphics[height=0.185\textwidth]{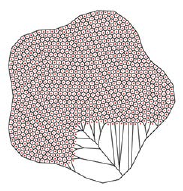}} \quad
\subfloat[]{\includegraphics[height=0.185\textwidth]{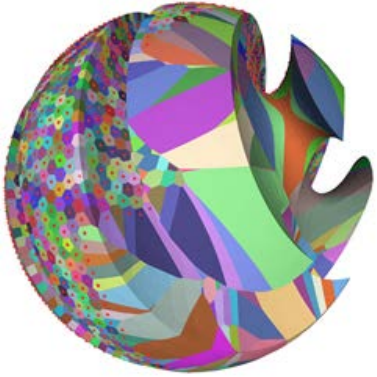}} \quad
\subfloat[]{\includegraphics[height=0.185\textwidth]{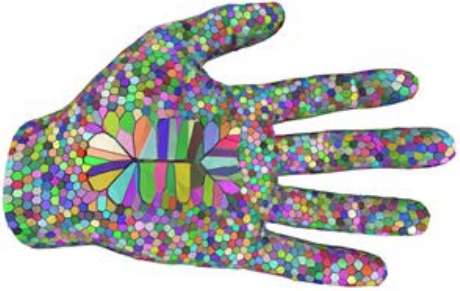}}

    \begin{tabular}{|c|ccccccccc|} \hline
        subfigure & (a) & (b) & (c) & (d) & (e) & (f) & (g) & (h) & (i)\\ \hline
        DT-based method~\cite{Aurenhammer1991-Voronoi, Yan2009-Remeshing,Yan2013-ClippedVoronoi} & \textbf{0.003} & 0.113 & 0.6 &  0.003 & 0.103 & 0.48 & \textbf{0.003} & 0.11 & 0.49 \\ 
        \textit{k}NN-based method~\cite{Levy2013-kNNVoronoi} & 0.004 & 0.065 & 0.684 &  0.003 & 0.05 & 0.465 & 0.006 & 0.267 & 1.296 \\ 
        our method (Algorithm~\ref{alg:serial}) & \textbf{0.003} & \textbf{0.05} & \textbf{0.462} &  \textbf{0.002} & \textbf{0.044} & \textbf{0.403} & \textbf{0.003} & \textbf{0.054} & \textbf{0.433} \\ \hline
    \end{tabular}

\caption{Clipped/restricted Voronoi diagrams of the given sites with different distributions. Top row: white noise; middle row: blue noise; bottom row: blue noise with partial sites removal; left column: 2D, $m=27, n=1$K; middle column: surface, $m=20.8$K, $n=5$K; right column: 3D, $m=20.5$K, $n=10$K. The runtimes (in seconds) for each result of the DT-based~\cite{Aurenhammer1991-Voronoi, Yan2009-Remeshing,Yan2013-ClippedVoronoi}, \textit{k}NN-based~\cite{Levy2013-kNNVoronoi} methods, and our method (Algorithm~\ref{alg:serial}) are provided in the inset table.}
\label{fig:voronoi-of-different-distributions}
\end{figure*}

\begin{figure*}[!t]
\centering
\subfloat[$m=180, n=1$K]{\includegraphics[height=0.198\textwidth]{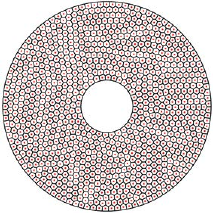}} \quad
\subfloat[$m=60.2$K, $n=5$K]{\includegraphics[height=0.198\textwidth]{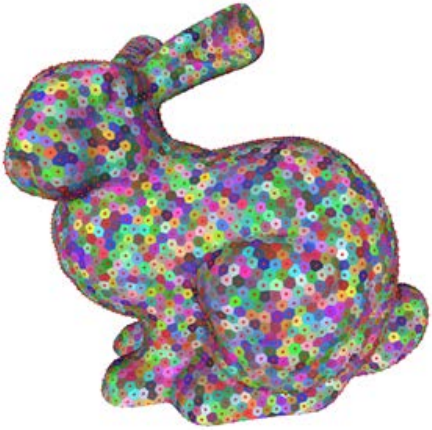}} \quad
\subfloat[$m=15.6$K, $n=10$K]{\includegraphics[height=0.198\textwidth]{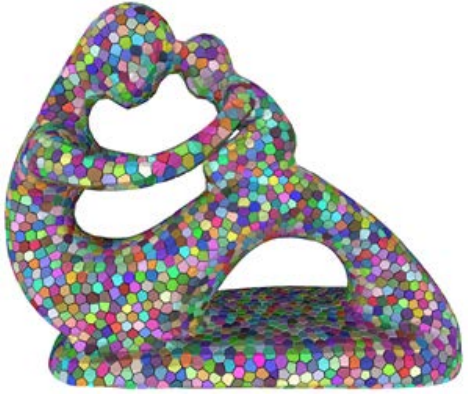}}
\caption{Another group of results, where the sites are with blue noise. The runtimes of the DT-based~\cite{Aurenhammer1991-Voronoi, Yan2009-Remeshing,Yan2013-ClippedVoronoi}, \textit{k}NN-based~\cite{Levy2013-kNNVoronoi} methods, and our method (Algorithm~\ref{alg:serial}) are (a) 2D: 0.005, 0.005, \textbf{0.004}; (b) surface: 0.125, 0.09, \textbf{0.08}; (c) 3D: 0.537, 0.51, \textbf{0.39} seconds, respectively.}
\label{fig:another-voronoi}
\end{figure*}

\section{Results and Applications}
\label{sec:results}

In this section, we present experimental results of our method and compare its performance with state-of-the-art methods in terms of both serial and parallel efficiency across 2D, surface, and 3D scenarios. Subsequently, we explore several applications of our method. All results are obtained on a desktop with the following specifications: an Intel Core i7-12700 CPU (2.1 GHz base clock), 16 GB RAM, and an NVIDIA RTX 3080 graphics card with 10 GB VRAM. For numerical values, K denotes thousand ($10^3$) and M denotes million ($10^6$). The following three types of site distributions will be mainly tested for comparing efficiency:

\begin{itemize}
    \item White noise: The sites are randomly sampled from the given domain.
    \item Blue noise: The sites are typically generated by centroidal Voronoi tessellation methods~\cite{Du1999-CVT, Liu2009_LBFGSCVT}.
    \item Perturbed grid: The sites are sampled with random perturbations from regular grid patterns.
\end{itemize}


\begin{figure*}[!t]
\centering

\begin{tabular}{c|cc|cc}
        2D & \multicolumn{2}{c|}{Surface} & \multicolumn{2}{c}{3D}\\
        rectangle ($m=1$) & sphere ($m=1.2$K) & bunny ($m=60.2$K) & ball ($m=1.9$K) & fertility ($m=15.6$K) \\
        \includegraphics[height=0.18\textwidth]{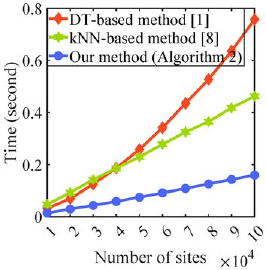} & \includegraphics[height=0.18\textwidth]{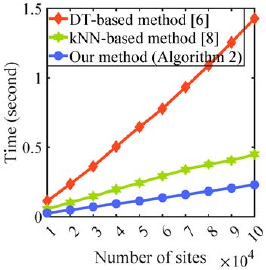} & \includegraphics[height=0.18\textwidth]{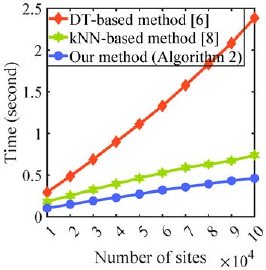} &  \includegraphics[height=0.18\textwidth]{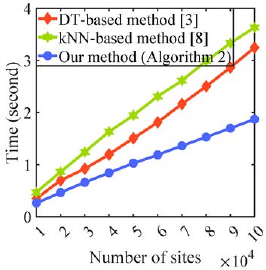} & \includegraphics[height=0.18\textwidth]{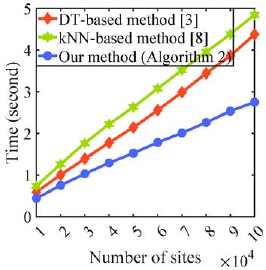} \\ 
        \includegraphics[height=0.18\textwidth]{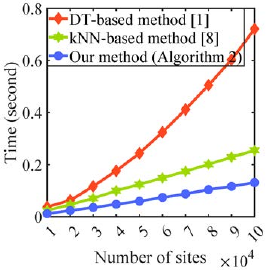} & \includegraphics[height=0.18\textwidth]{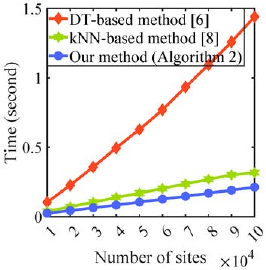} & \includegraphics[height=0.18\textwidth]{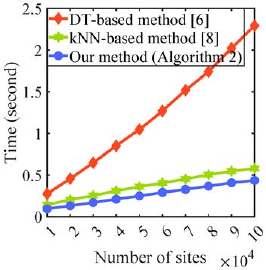} &  \includegraphics[height=0.18\textwidth]{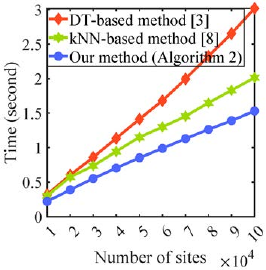} & \includegraphics[height=0.18\textwidth]{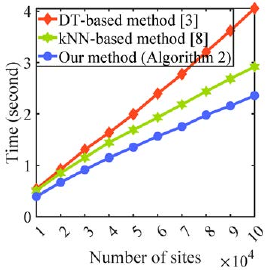}
    \end{tabular}

\caption{Runtime curves of the DT-based~\cite{Aurenhammer1991-Voronoi, Yan2009-Remeshing,Yan2013-ClippedVoronoi}, \textit{k}NN-based~\cite{Levy2013-kNNVoronoi} methods, and our method (Algorithm~\ref{alg:serial}) for computing clipped or restricted Voronoi diagrams across different domains with varying numbers of sites. The curves compare performance under white noise (top) and blue noise (bottom), respectively.}
\label{fig:time-vnb}
\end{figure*}

\subsection{Serial performance}

We compare our method with the DT-based~\cite{Aurenhammer1991-Voronoi, Yan2009-Remeshing,Yan2013-ClippedVoronoi} and \textit{k}NN-based~\cite{Levy2013-kNNVoronoi} methods while ensuring exact results. All these methods are implemented in C++, with the Delaunay triangulation for the DT-based method constructed using the CGAL library~\cite{cgal:eb-22b}. For the \textit{k}NN-based method, the nearest neighbors list of each site is enlarged independently if the security radius is not met. 

Fig.~\ref{fig:voronoi-of-different-distributions} presents clipped or restricted Voronoi diagrams on 2D, surface, and 3D domains. The rows correspond to different site distributions: white noise (top), blue noise (middle), and blue noise with partial sites removal (bottom). The runtimes of three methods for generating each result are given in the inset table. It is clear that the efficiency of the \textit{k}NN-based method fluctuates more intensely than the DT-based method and ours. When a site dominates points far away from it, the \textit{k}NN-based method has to launch many more clippings to meet the security radius for generating its cell, even though most of the clippings are invalid. Thus, it performs even worse than the DT-based method in such cases, see Fig.~\ref{fig:voronoi-of-different-distributions}(c, g, h, i). In contrast, our method remains more stable and consumes much less time than the others across all site distributions. Another group of results are given in Fig.~\ref{fig:another-voronoi}, where our method still costs the least.

The runtime curves of three methods against the number of sites are plotted in Fig.~\ref{fig:time-vnb}. The tested domains include: a rectangle (2D, $m=1$), two triangular surface meshes (sphere, $m=1.2$K and bunny, $m=60.2$K) and two tetrahedral meshes (ball, $m=1.9$K and fertility, $m=15.6$K). The site distributions correspond to white noise (top) and blue noise (bottom), respectively. With increasing numbers of sites, the computational cost of Delaunay triangulation construction in the DT-based method grows significantly, resulting in the lowest efficiency in most cases. The \textit{k}NN-based method performs the worst in the 3D case with white noise, while our method achieves the highest efficiency across all cases.


\begin{figure}[!t]
\centering

\begin{tabular}{c|c}
        Surface (sphere) & 3D (ball)\\
        \includegraphics[height=0.18\textwidth]{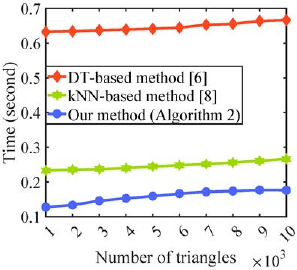} & \includegraphics[height=0.18\textwidth]{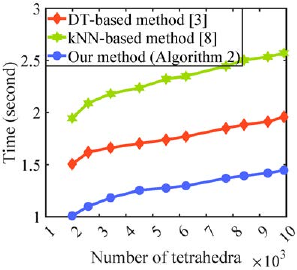} \\ 
        \includegraphics[height=0.18\textwidth]{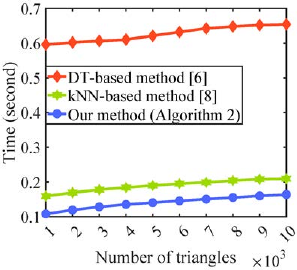} & \includegraphics[height=0.18\textwidth]{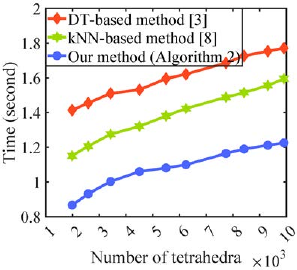}
    \end{tabular}

\caption{Runtime curves of the DT-based, \textit{k}NN-based methods, and our method for computing clipped or restricted Voronoi diagrams with varying numbers of domain simplices, where the number of sites is fixed at $n = 50$K. The site distributions are white noise (top) and blue noise (bottom).}
\label{fig:region-time}
\end{figure}

Furthermore, by fixing the number of sites ($n=50$K), we compare the performance of three methods against the number of domain simplices, their runtime curves are presented in Fig.~\ref{fig:region-time}, where the domains are a triangular mesh (left) and a tetrahedral mesh (right), and the site distributions are white noise (top) and blue noise (bottom), respectively. This comparison reaffirms the superior performance of our method, since these curves exhibit a similar trend to that in Fig.~\ref{fig:time-vnb}: the \textit{k}NN-based method is the most time-consuming in the 3D case with randomly generated sites, while the DT-based method performs worst in other scenarios. In contrast, our method consistently consumes the least time.

\begin{figure}[t]
  \centering
  \subfloat[white noise]{\includegraphics[height=0.16\textwidth]{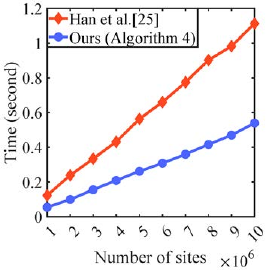}} \quad
  \subfloat[blue noise]{\includegraphics[height=0.16\textwidth]{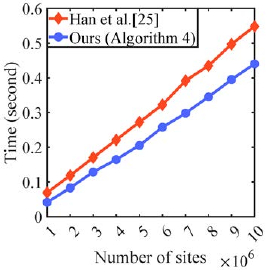}}
  \caption{Runtime comparison between Han et al.~\cite{Han2017-GPURVD} and Algorithm 4 on a surface domain (bunny, $m = 60.2$K) with varying site numbers. The sites are with white noise (a) and blue noise (b), respectively.}
  \label{fig:GPU-time-bunny}
\end{figure}

\begin{figure}[t]
  \centering
  \subfloat[white noise]{\includegraphics[height=0.16\textwidth]{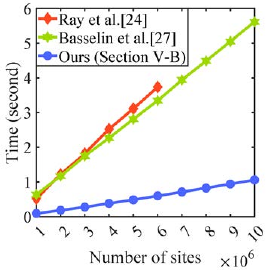}}
  \subfloat[perturbed grid]{\includegraphics[height=0.16\textwidth]{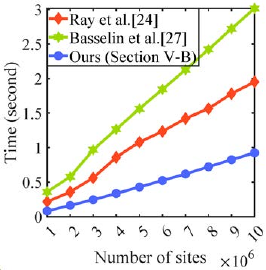}}
  \subfloat[blue noise]{\includegraphics[height=0.16\textwidth]{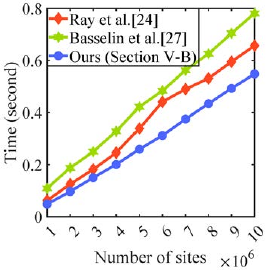}}
  \caption{Runtime comparison between Ray et al.~\cite{Ray2018-MeshlessVoronoi}, Basselin et al.~\cite{Basselin2021-GPUknnRPD} and our parallel algorithm (Section \ref{sec:parallel-regular}) on a simple 3D domain (cube, $m = 1$) with varying site numbers. The sites are with white noise (a), perturbed grid (b) and blue noise (c), respectively.}
  \label{fig:GPU-time-cube}
\end{figure}

\begin{figure}[t]
  \vspace{-10pt}
  \centering
  \subfloat[white noise]{\includegraphics[height=0.16\textwidth]{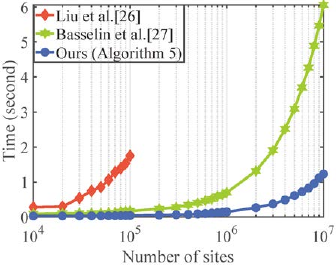}}
  \subfloat[blue noise]{\includegraphics[height=0.16\textwidth]{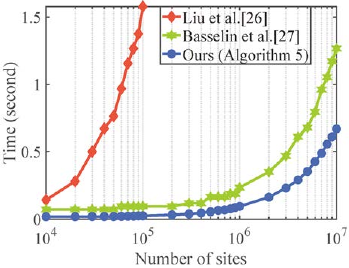}}
  \caption{Runtime comparison between Liu et al.~\cite{Liu2022-GPU3DVoronoi}, Basselin et al.~\cite{Basselin2021-GPUknnRPD} and Algorithm 5 on a complex 3D domain (fertility, $m = 15.6$K) with varying site numbers. The sites are with white noise (a) and blue noise (b), respectively.}\label{fig:GPU-time-fertility}
\end{figure}

\begin{figure}[t]
  \vspace{-2pt}
  \centering
  \subfloat[white noise]{\includegraphics[height=0.16\textwidth]{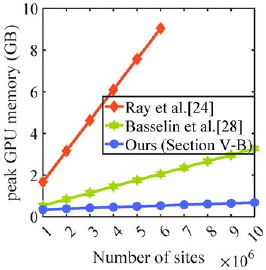}}
  \subfloat[perturbed grid]{\includegraphics[height=0.16\textwidth]{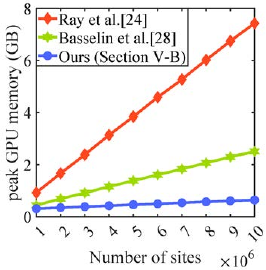}}
  \subfloat[blue noise]{\includegraphics[height=0.16\textwidth]{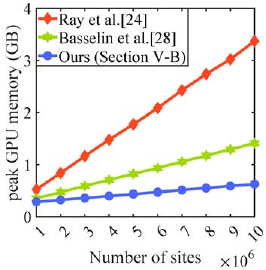}}
  \caption{Comparison of peak GPU memory usage between Ray et al.~\cite{Ray2018-MeshlessVoronoi}, Basselin et al.~\cite{Basselin2021-GPUknnRPD} and our parallel algorithm (Section \ref{sec:parallel-regular}) on a simple 3D domain (cube, $m = 1$) with varying site numbers. The sites are with white noise (a), perturbed grid (b) and blue noise (c), respectively.}
  \label{fig:GPU-memory-cube}
\end{figure}

\subsection{Parallel performance}

The parallel version of our method still ensures exact results and rapidly delivers outcomes under varying site distributions. To evaluate the performance of our parallel algorithm, we compare it with the state-of-the-art method, namely the \textit{k}NN-based method. Several parallel versions of the \textit{k}NN-based method have been developed for different domains: surface meshes~\cite{Han2017-GPURVD}, simple 3D volumes~\cite{Ray2018-MeshlessVoronoi}, and complex 3D volumes~\cite{Liu2022-GPU3DVoronoi, Basselin2021-GPUknnRPD}. We implement our parallel algorithm and the algorithm of Han et al.~\cite{Han2017-GPURVD} using CUDA, while the implementations of other \textit{k}NN-based versions are obtained from the following sources: Ray et al.~\cite{Ray2018-MeshlessVoronoi}\footnote{https://dl.acm.org/doi/10.1145/3272127.3275092}, Liu et al.~\cite{Liu2022-GPU3DVoronoi}\footnote{https://github.com/xh-liu-tech/3D-Voronoi-GPU}, and Basselin et al.~ \cite{Basselin2021-GPUknnRPD}\footnote{https://github.com/basselin7u/GPU-Restricted-Power-Diagrams}. All these algorithms are tested using 32-bit floating-point numbers.

Given a triangular surface mesh (bunny, $m=60.2$K), a cube ($m=1$), and a tetrahedral mesh (fertility, $m=15.6$K), each with sites following different distributions and scales, the runtime curves of our parallel algorithm and the parallel implementations of the \textit{k}NN-based method are shown in Figs.~\ref{fig:GPU-time-bunny}$\sim$\ref{fig:GPU-time-fertility}. Each reported runtime covers cells construction and centroids computation, excluding GPU memory allocation and data transfer. Note that in Fig.~\ref{fig:GPU-time-cube}, for the white noise distribution, the algorithm of Ray et al.~\cite{Ray2018-MeshlessVoronoi} requires GPU memory beyond the test machine's capacity when the number of sites exceeds 7M. Hence, for the white noise case, the number of sites tested with their algorithm is limited to 6M. Similarly, in Fig.~\ref{fig:GPU-time-fertility}, the algorithm of Liu et al.~\cite{Liu2022-GPU3DVoronoi} is tested with no more than 100K sites. These runtime curves demonstrate that our parallel algorithm consistently outperforms the \textit{k}NN-based parallel algorithms across all site distributions, with a clear speed advantage observed for white noise and perturbed grid distributions.

Our parallel algorithm not only improves efficiency but also reduces GPU memory requirements. The parallel implementations of the \textit{k}NN-based method must store the \textit{k}-nearest neighbors for each site, resulting in substantial memory overhead. Although Basselin et al.~ \cite{Basselin2021-GPUknnRPD} introduced a multi-pass strategy to alleviate this demand, memory consumption remains high. Our algorithm, however, only requires a structure for nearest-neighbor searches, leading to markedly reduced GPU memory requirements. Fig.~\ref{fig:GPU-memory-cube} compares the peak GPU memory usage of our parallel algorithm with those of Ray et al.~\cite{Ray2018-MeshlessVoronoi} and Basselin et al.~ \cite{Basselin2021-GPUknnRPD} on a cube under varying site distributions and scales. The results show that our algorithm consumes significantly less memory across all tested cases.

\subsection{Computation of power diagrams}

The power diagram (weighted Voronoi diagram)~\cite{Aurenhammer1987-Power} is an extension of the Voronoi diagram, and computing power diagrams as fast as possible is important to downstream applications. Given a set of weighted sites $\left \{ (\mathbf{v}_i, w_i) \right \}_{i=1}^n$, a power diagram can be traditionally computed by using the regular triangulation (RT) based method~\cite{Edelsbrunner1996-RT}. It has been proven that any power diagram equals to a restricted Voronoi diagram in higher one dimensional space~\cite{Xiao2023-knnpower}, so that the Voronoi generation methods can be directly used after the given weighted sites are lifted as $\left \{ \mathbf{p}_i = (\mathbf{v}_i, \sqrt{\eta - w_i})\right \}_{i=1}^n$, where $\eta \geq w_{\text{max}} = \max \left \{w_1,\dots, w_n \right \}$. The \textit{k}NN-based method has been extended to compute power diagrams~\cite{Xiao2023-knnpower}, and we apply our method to the computation in this section.

\begin{figure*}[!t]
\centering
\subfloat[$m=1,n=1$K]{\includegraphics[height=0.22\textwidth]{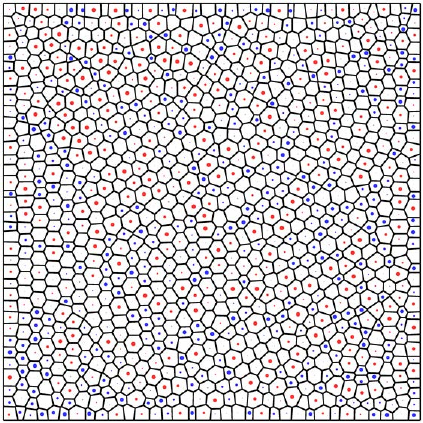}} \quad \quad
\subfloat[$m=149.5$K, $n=5$K]{\includegraphics[height=0.22\textwidth]{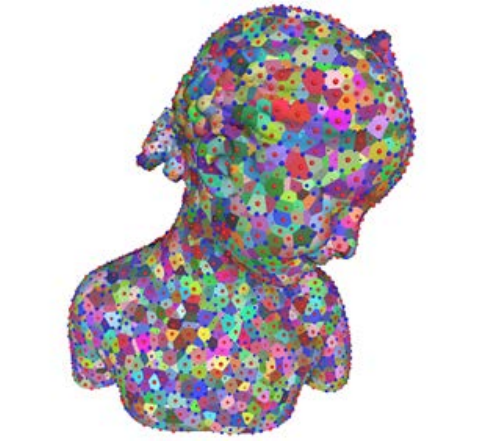}} \quad
\subfloat[$m=24.5$K, $n=10$K]{\includegraphics[height=0.22\textwidth]{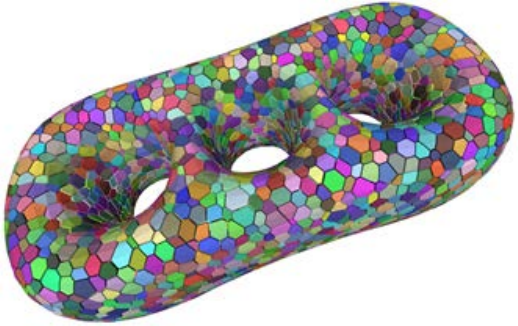}}
\caption{Our method can be directly applied to the computation of power diagrams after the given weighted sites are lifted. The given weighted sites are with blue noise locations and random weights. The sites with positive and negative weights are rendered as red and blue disks/balls, respectively, and the absolute value of each weight is proportional to the size of corresponding disk/ball. To obtain these results, the runtimes of the RT-based, \textit{k}NN-based methods and ours are (a) 2D: 0.003, \textbf{0.002}, \textbf{0.002}, (b) surface: 0.326, 0.166, \textbf{0.145}, (c) 3D: 0.552, 0.526, \textbf{0.46} seconds, respectively.}
\label{fig:power-computation}
\end{figure*}

Fig.~\ref{fig:power-computation} shows three examples of clipped or restricted power diagrams generated by our method, where the given weighted sites are with blue noise locations and random weights, the weights are in $[-|\mathcal{M}| / n, |\mathcal{M}| / n]$ and $|\mathcal{M}|$ is the total area/volume of the given domain. We use the RT-based and \textit{k}NN-based methods and our method to generate the same results and compare their running time. A similar conclusion is conducted that our method consumes the least computational time for each case. It is worth pointing out that because the weighted sites are lifted to a higher dimensional space, our method may consume more time on the nearest neighbor searches when computing power diagrams compared to computing Voronoi diagrams.

It has been shown that the parameter $\eta$ influences the efficiency of the \textit{k}NN-based method because it determines whether the security radius can be met early. Thus, the minimal $\eta$ (i.e., $\eta = w_{\text{max}}$) is optimal for the \textit{k}NN-based method~\cite{Xiao2023-knnpower}. Similarly, $\eta$ also has a high impact on our method's efficiency. A larger $\eta$ generates lifted sites far away from the domain, causing our method to traverse more grids during each nearest neighbor search and thereby increasing computational time. Hence, we suggest using the minimal $\eta$ (i.e., $\eta = w_{\text{max}}$) for lifting weighted sites when computing power diagrams with our method.

\subsection{Centroidal Voronoi tessellations}

\begin{figure*}[!t]
\centering

\begin{tabular}{cc|cc}
        \multicolumn{2}{c|}{2D CVT} & \multicolumn{2}{c}{3D CVT}\\
        \includegraphics[height=0.22\textwidth]{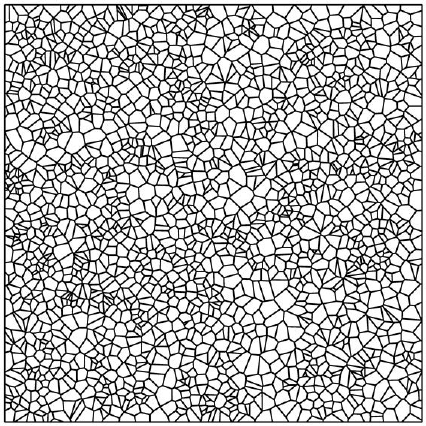} & \includegraphics[height=0.22\textwidth]{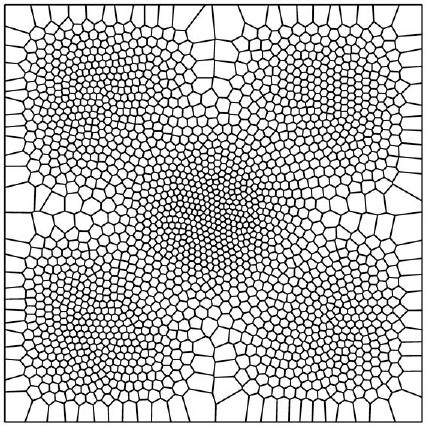} &  
        \includegraphics[height=0.22\textwidth]{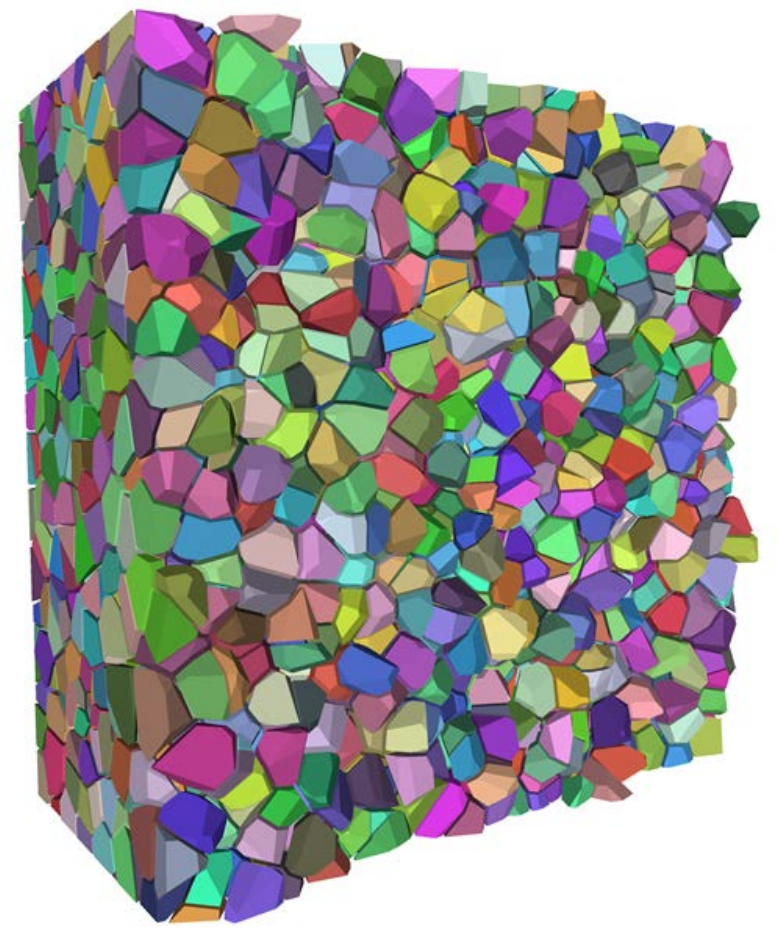} & \includegraphics[height=0.22\textwidth]{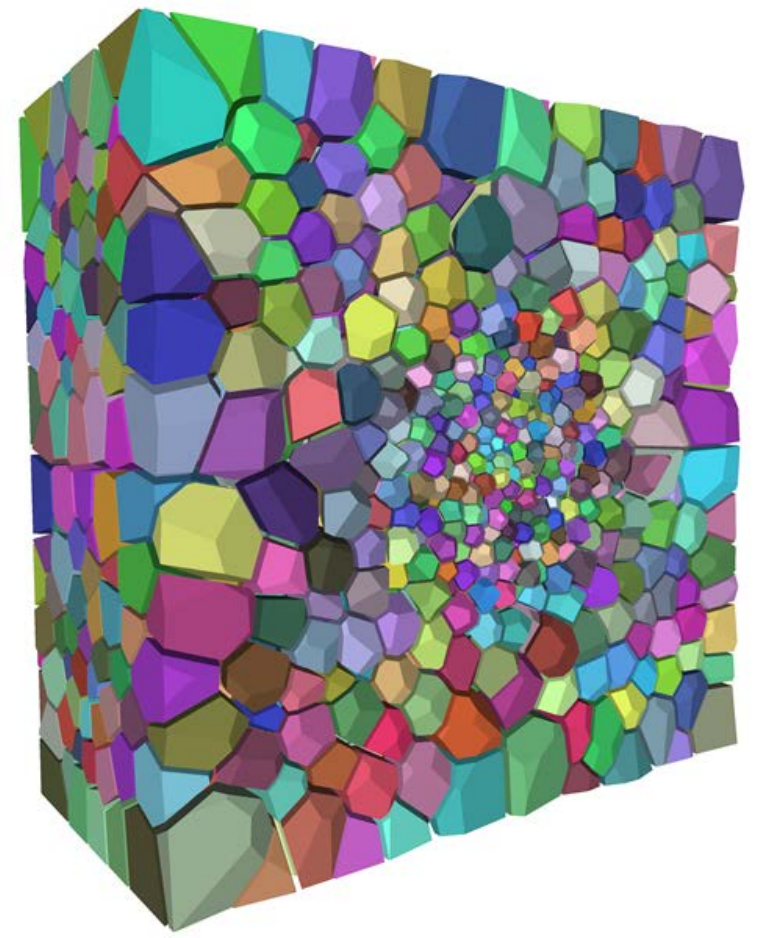} \\
        \text{initialization} & \text{result} & \text{initialization} & \text{result}
\end{tabular}

\caption{2D and 3D CVT results generated by the Lloyd's method with 200 iterations (2D: $n=2$K, 3D: $n=10$K). For computing Voronoi diagrams in generating the CVT results, our method saves 34\% (2D) / 42.4\% (3D) of time compared to the DT-based method~\cite{Aurenhammer1991-Voronoi,Yan2013-ClippedVoronoi}, and 46.8\% (2D) / 73.5\% (3D) of that compared to the \textit{k}NN-based method~\cite{Levy2013-kNNVoronoi}. For the sites with $z > 0$ in the 3D case, their cells are not shown.}
\label{fig:CVT}
\end{figure*}

The centroidal Voronoi tessellation (CVT) is a special Voronoi diagram in which each site coincides with the centroid of its Voronoi cell. It also can be defined as the tessellation when the following energy function is minimized~\cite{Du1999-CVT,Liu2009_LBFGSCVT}:
\begin{equation}
    \label{equ:cvt}
    \mathcal{E}(\left \{ \mathbf{v}_i \right \}_{i=1}^n) = \sum_{i=1}^n \int_{\Omega_i} \rho(\mathbf{x}) \| \mathbf{x} - \mathbf{v}_i \|^2 d \mathbf{x},
\end{equation}
where $\rho(\mathbf{x})$ is the density function. Here, the Lloyd's method~\cite{Du1999-CVT} is adopted to minimize equation (\ref{equ:cvt}). In each iteration, we compute centroids of Voronoi cells as the new sites and update the Voronoi diagram. Fig.~\ref{fig:CVT} shows 2D and 3D CVT results after 200 iterations starting from random initialization, where $\rho(\mathbf{x}) = e^{-20 \| \mathbf{x} \|^2} + 0.05\prod_{k=1}^{d}\sin^2(\pi \mathbf{x}(k))$, $n=2$K and 10K in the 2D and 3D cases, respectively. The total time costs for computing centroids are 1.21 (2D) and 370.15 (3D) seconds. For the computation of Voronoi diagrams, we tested the DT-based~\cite{Aurenhammer1991-Voronoi,Yan2013-ClippedVoronoi}, \textit{k}NN-based~\cite{Levy2013-kNNVoronoi} methods and ours, which cost in total 0.779, 0.966, and 0.514 seconds in the 2D case, and 39.139, 85.298, and 22.535 seconds in the 3D case. Our method saves 34\% (2D) / 42.4\% (3D) of time compared to the DT-based method, and 46.8\% (2D) / 73.5\% (3D) of that compared to the \textit{k}NN-based method, revealing the efficiency advantage of our method. 

We also perform the Lloyd's method on GPUs. Starting from Fig.~\ref{fig:teaser}(a) with 5M sites distributed in white noise, the CVT result after 100 iterations is shown in Fig.~\ref{fig:teaser}(b), where $\rho(\mathbf{x}) = 1$. Our parallel algorithm and the \textit{k}NN-based parallel algorithm~\cite{Ray2018-MeshlessVoronoi} are both tested for generating Voronoi cells in this example. The result obtained using our parallel algorithm takes only 31.07 seconds, while it requires 54.8 seconds if the \textit{k}NN-based parallel algorithm~\cite{Ray2018-MeshlessVoronoi} is applied.

\section{Conclusion and Future Work}
\label{sec:conclusion}

In this paper, we propose an efficient method for computing clipped or restricted Voronoi diagrams. The key distinction between our method and existing approaches lies in the computation of each simplex-cell intersection. A clipping will be launched only if an edge of the intermediate intersection has endpoints being respectively inside and outside a target Voronoi cell. The corresponding clipping plane is rapidly identified through our presented edge-based bisector search scheme. Consequently, all the clippings in our method are valid to final results, demonstrating a great advantage over existing approaches. Additionally, we introduce a parallel implementation of our method, further accelerating Voronoi diagram computations. Extensive experimental evaluations confirm that our method outperforms state-of-the-art ones in both serial and parallel efficiency. It is worth exploring broader applications of our method in future work.

\bibliographystyle{IEEEtran}
\bibliography{refs}

\end{document}